\documentclass[lettersize,journal]{IEEEtran}
\usepackage{subfigure}
\usepackage{comment}
\usepackage{placeins}
\usepackage{tabularx}
\usepackage{booktabs}
\usepackage{float}
\usepackage{color}
\usepackage{cite}
\usepackage{citesort}
\usepackage{amsmath,amssymb,amsfonts}
\usepackage{algorithmic}
\usepackage{graphicx}
\usepackage{textcomp}
\usepackage{placeins}
\usepackage{xcolor}
\usepackage{color}
\usepackage{lettrine}
\usepackage{citesort}
\usepackage{cite}
\usepackage{amsfonts}
\usepackage{amssymb}
\usepackage{amsmath}
\usepackage{mathrsfs}
\usepackage{cite}
\usepackage{balance}
\usepackage{booktabs}
\usepackage{fancybox}
\usepackage{bm}
\usepackage{extarrows}
\usepackage{algorithm}
\usepackage{multirow}
\usepackage{array}

\def\BibTeX{{\rm B\kern-.05em{\sc i\kern-.025em b}\kern-.08em
    T\kern-.1667em\lower.7ex\hbox{E}\kern-.125emX}}
    
\begin{document}
\UseRawInputEncoding


\title{Nonreciprocal RIS-Aided Covert Channel Reciprocity Attacks and Countermeasures}

\author
{\IEEEauthorblockN{Haoyu~Wang, Jiawei~Hu, Jiaqi Xu,~\IEEEmembership{Member, IEEE}, Ying Ju,~\IEEEmembership{Member, IEEE},\\and A. Lee~Swindlehurst,~\IEEEmembership{Fellow, IEEE}} 

\thanks{H. Wang, J. Xu, and A. L. Swindlehurst are with the Center for Pervasive Communications and Computing, University of California at Irvine, Irvine, CA 92697 USA (e-mail: haoyuw30@uci.edu;  xu.jiaqi@uci.edu; swindle@uci.edu ).}
\thanks{J. Hu and Y. Ju are with the School of Telecommunications Engineering, Xidian University, Xi'an 710071, China (e-mail: jiaweihu27@xidian.edu.cn; juying@xidian.edu.cn).}

}


\maketitle

\begin{abstract}
Reconfigurable intelligent surface (RIS) technology enhances wireless communication performance, but it also introduces new vulnerabilities that can be exploited by adversaries. This paper investigates channel reciprocity attack (CRACK) threats in multi-antenna wireless systems operating in time-division duplexing mode using a physically consistent non-reciprocal RIS (NR-RIS) model. CRACK can degrade communication rate and facilitate passive eavesdropping behavior by distorting the downlink precoding, without requiring any additional signal transmission or channel state information (CSI). Unlike conventional RIS jamming strategies, the NR-RIS does not need synchronization with the legitimate system and thus can operate with slow or fixed configurations to implement CRACK, obscuring the distinction between the direct and RIS-induced channels and thereby complicating corresponding defensive precoding designs. To counter the CRACK threat posed by NR-RIS, we develop ``SecureCoder,'' a deep reinforcement learning-based framework that can mitigate CRACK and determine an improved downlink precoder matrix using the estimated uplink CSI and rate feedback from the users. Simulation results demonstrate the severe performance degradation caused by NR-RIS CRACK and validate the effectiveness of SecureCoder in improving both throughput and reducing security threats, thereby enhancing system robustness.
\end{abstract}

\begin{IEEEkeywords}
Channel Reciprocity, Passive Attack, Reconfigurable Intelligent Surfaces, Eavesdropping, TDD system.
\end{IEEEkeywords}

\IEEEpeerreviewmaketitle

\section{INTRODUCTION}\label{s1}

\IEEEPARstart{C}{ommunication} 
security is an essential and challenging problem in wireless networks due to the open nature of wireless channels, making them vulnerable to various threats~\cite{7467419}. Besides traditional encryption methods in the application layer, physical layer security (PLS) techniques have emerged as potential solutions in next-generation networks~\cite{6739367,7081071}. Concurrently, reconfigurable intelligent surfaces (RIS) have been introduced as a promising tool for various applications in future wireless networks, including enhancing communication security \cite{10409564,10188924}. An RIS comprises an array of passive elements whose reflection coefficients can be dynamically configured through control signals. With sufficient channel state information (CSI), an RIS can create "virtual" line-of-sight (LoS) paths around obstacles, enhance the desired signal power by constructively superimposing reflected signals with direct paths, or mitigate interference through destructive combination with the direct path~\cite{9475160,9086766,9424177}. Many studies have shown that RIS can effectively mitigate physical layer threats in wireless systems, including jamming and eavesdropping attacks, using similar principles~\cite{9198898,10143420,9779086,9439833,10143983,9501003,9963962}. 
However, RIS technology may also be exploited for malicious purposes if controlled by an adversary. Such attacks could involve compromising the controller of a legitimate RIS resulting in unintended operation \cite{9941040}, deploying an RIS to enhance signal leakage towards an eavesdropper, or increasing the jamming power at specific receivers \cite{wei2023metasurface,9789438,9605003,10073942}. Another type of attack involves manipulating the RIS to destructively combine the reflected signal with the direct path signal, thereby reducing the signal-to-noise ratio (SNR) or completely canceling the signal at the receiver \cite{9112252,10302337,10516473}. These types of attacks generally require the adversary to possess CSI for the targeted links, which can be challenging to achieve in practice.

\subsection{RIS-Aided Passive Jamming Attacks on TDD Systems}
Recent studies have proposed an RIS-based attack strategy that does not require explicit channel state information (CSI) in order to compromise time-division duplex (TDD) multiuser multiple-input single-output (MU-MISO) systems, where downlink precoding relies on uplink pilot-based CSI estimation. In~\cite{10081025}, a pilot contamination attack is introduced in which an adversarial RIS applies one set of random phase shifts during uplink training and a different set during downlink transmission. This variation in the channel results in a mismatched precoder, causing unanticipated multiuser interference and degrading downlink performance. A variation in~\cite{10149173} considers an adversarial RIS that remains silent during uplink estimation but employs rapidly time-varying phase shifts in the downlink. Similarly, \cite{10682037} explores attacks where the RIS randomly changes its configuration once during uplink estimation and rapidly varies during downlink transmission. In~\cite{10402016}, another variant is proposed for MU-MIMO systems, where the RIS remains silent during uplink transmissions and then optimizes its phase in real-time based on instantaneous knowledge of the RIS-induced links during each downlink period. These techniques are collectively referred to as ``passive jamming" attacks, as they degrade system performance without active transmissions, making them difficult to detect.

A notable limitation of the methods proposed in~\cite{10081025,10149173,10424421,10682037} is that the malicious RIS must remain synchronized with both the uplink training and downlink data transmission phases of the targeted MU-MISO network. Furthermore, the rapid variations in CSI at rates significantly exceeding the normal channel coherence time can be readily detected by legitimate systems, thereby undermining the covertness of the attack. In~\cite{10445725}, a method referred to as CRACK for achieving 
a Channel Reciprocal AttaCK was presented that eliminates these drawbacks, while still requiring no CSI and nor active transmissions. The method of \cite{10445725} relies on the use of a so-called ``non-diagonal'' RIS \cite{9737373,10302331} (ND-RIS), in which the signal received by one RIS element can be reflected through another RIS element, after proper configuration. From an attacker's perspective, the key benefit of this type of RIS is that it can naturally lead to non-reciprocal propagation since the phase-shift matrix i non-symmetric. Thus, the ND-RIS idea inherently and passively makes the uplink channel different from the downlink channel without requiring any CSI nor synchronization with the system, and without continually and rapidly changing its configuration.

\subsection{Related Countermeasures}
Recent work has focused on attempting to mitigate some of the above passive jamming attacks.
In~\cite{10402016}, the authors assume that the base station (BS) can estimate the RIS-induced channels in real time, thus minimizing signal leakage into these channels via the downlink precoding design. The anti-jamming precoding approach of~\cite{10424421} counteracts the ``channel aging" attacks of \cite{10081025,10149173} by leveraging statistical characteristics introduced by the rapid variations in RIS-induced channels. The work in~\cite{10500386,10247269} explores the impact of an RIS reciprocity attack on TDD-based physical layer key generation under the assumption of rapidly varying RIS-induced channels. A corresponding defensive strategy is also proposed, wherein fast-varying RIS-induced paths are identified and eliminated through a process referred to as ``contaminated path removal." Subsequently, the legitimate transmitter and receiver negotiate an appropriate path set for secure key generation. This countermeasure is potentially effective against the methods discussed 
in~\cite{10424421,10149173,10682037}, which rely on RIS configurations that change significantly faster than the normal channel coherence time. However, none of the above methods is effective against the CRACK approach, since in principle it does not require any time varying adjustment. Except for its non-reciprocal behavior, it functions analogously to naturally occurring time-invariant scattering in the environment. Since signals reflected by a CRACK-enabled RIS are generally indistinguishable from natural multipath components in the environment, countering CRACK presents a significant challenge. 

\subsection{Motivations and Contributions}
While CRACK poses a critical threat to TDD-based systems, the ND-RIS model in \cite{10445725} is challenging to implement in practice because it does not directly follow from physically consistent assumptions. Some initial work on physically realizable non-reciprocal RIS (NR-RIS) has been presented in~\cite{10694006,nrris,li2024non}. In particular, \cite{nrris} demonstrated that there is no loss of attack performance when restricting attention to physically consistent models compared to the idealized model of \cite{10445725}. For either modeling assumption, no effective countermeasures have been proposed to mitigate CRACK. 

In this paper, we investigate CRACK threats introduced by NR-RIS using a physically consistent model in TDD communication systems, where the NR-RIS can not only dramatically reduce the throughput but also strengthen the eavesdropping ability of an adversary. Although CRACK does not require knowledge of the legitimate system, when such information is available the adversary can optimize the NR-RIS to achieve a more severe attack. To mitigate these threats, we propose an intelligent anti-CRACK precoding approach based on deep reinforcement learning (DRL) that we refer to as ``SecureCoder.'' Once trained, SecureCoder provides robust downlink transmissions operating solely based on the estimated overall uplink CSI at the BS. The key contributions of this work are summarized as follows:
\begin{itemize}
\item We introduce a novel NR-RIS model using multiport network analysis to investigate the impact of CRACK on TDD-based multi-user wireless systems, where the NR-RIS scattering matrix and induced channels follow physical-consistent models. When an NR-RIS reflects sufficient user signal energy to the BS, it covertly disrupts channel reciprocity and thus leads to distorted downlink precoding at the BS, significantly deteriorating the system performance. This passive method operates without transmit power, does not rely on CSI from the legitimate system, and requires no synchronization with the target network. In addition, the NR-RIS can remain static without frequent state changes.
\item We explore three potential CRACK scenarios including blind CRACK, CRACK-aided eavesdropping, and knowledge-driven CRACK, each of which is difficult to detect or mitigate. We present numerical results to demonstrate that these proposed attack strategies can induce substantial throughput degradation and information leakage in TDD systems without prior knowledge about the legitimate network. However, we show that if such knowledge is available, the attack effect can be dramatically enhanced.
\item To address the complexity that arises from allowing arbitrary element pairing in large-scale NR-RIS, we adopt a block-based architecture, where elements are flexibly paired within independent blocks of the NR-RIS elements. Simulations reveal that this design achieves attack performance comparable to the ideal case with arbitrary interconnections, and outperforms both the ND-RIS model in~\cite{10445725} and the passive jamming approaches in~\cite{10081025,10149173,10424421}.
\item We propose the SecureCoder technique based on DRL to partially counteract the impact of NR-RIS CRACK. SecureCoder implements an enhanced proximal policy optimization algorithm that enables the BS to learn an effective precoding strategy in the non-reciprocal environment created by the NR-RIS. Once trained, SecureCoder only requires the estimated uplink CSI to determine the downlink precoding matrix. Simulations show that, with proper training, SecureCoder realizes improved downlink transmission under both NR-RIS CRACK and diagonal-RIS passive jamming scenarios such as~\cite{10149173}.
\end{itemize}

The remainder of the paper is organized as follows. In Section~\uppercase\expandafter{\romannumeral2}, the operation of a TDD MU-MISO system under NR-RIS CRACK is described using a physically consistent multiport network model.  Section~\uppercase\expandafter{\romannumeral3} explores three different covert CRACK strategies and summarizes their characteristics and optimization objectives.
Section~\uppercase\expandafter{\romannumeral4} introduces the DRL-based ``SecureCoder'' framework, and Section~\uppercase\expandafter{\romannumeral5} presents several representative simulation results. Finally, conclusions are drawn in Section~\uppercase\expandafter{\romannumeral6}.

\textit{Notation}: Vectors and matrices are denoted by boldface lower and upper case letters, respectively. The operators $(\cdot)^T$, $(\cdot)^H$, and $(\cdot)^*$ represent the transpose, Hermitian transpose, and conjugate operations, respectively. The magnitude of the complex scalar $x$, the $2$-norm of vector $\mathbf{x}$, and the Frobenius norm of matrix $\mathbf{X}$ are respectively denoted by $|x|$, $\Vert \mathbf{x}\Vert$, and $\Vert \mathbf{X}\Vert_{\rm F}$. The matrix ${\rm blkdiag}\left\{\mathbf{A}_1,\mathbf{A}_2,\dots \mathbf{A}_N\right\}$ denotes a block-diagonal matrix whose diagonal blocks are given by the submatrix arguments.

\section{System Description}\label{s2}

\begin{figure}[!t]
\begin{center}
\includegraphics[width=3 in]{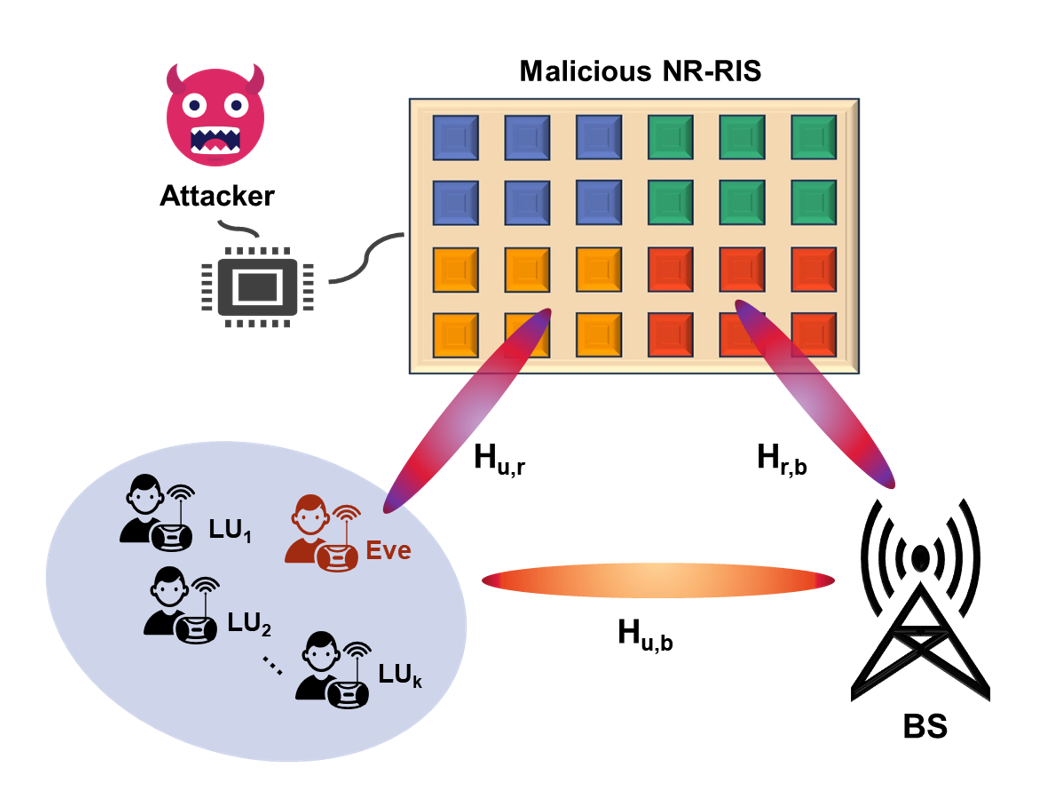}
\end{center}
\vspace{-0.08in}
\caption{Illustration of a TDD MU-MISO system with a malicious NR-RIS that destroys channel reciprocity.}\label{f1}
\vspace{-0.13in}
\end{figure}
Fig. \ref{f1} illustrates the considered MU-MISO system, where an $M$-antenna BS communicates with $K < M$ single-antenna legitimate users ($\rm{LUs}$) denoted as $\rm{LU_1}, \rm{LU_2},...,\rm{LU_K}$. Meanwhile, an adversarial NR-RIS with $N$ elements is present and, unknown to the BS, affects the propagation of signals to/from the BS. Fig.~\ref{f1} also shows that an eavesdropper may also be present in the network; this possibility will be discussed later. The terms $\mathbf{h}_{k,r}\in \mathbb{C}^{N \times 1}, \mathbf{H}_{r,b}\in \mathbb{C}^{M \times N}$ and $\mathbf{h}_{k,b}\in \mathbb{C}^{M \times 1}$, respectively denote the flat fading channels from $\rm{LU_k}$ to the NR-RIS, from the NR-RIS to the BS, and from $\rm{LU_k}$ to the BS. The channel $\mathbf{H}_{up}\in \mathbb{C}^{M \times K}$ is the overall cascaded channel between the BS and users, including the NR-RIS, while $\mathbf{H}_{down}=\mathbf{H}_{up}^T$ is the channel that the BS {\em assumes} for the downlink channel under reciprocity.

\subsection{TDD MU-MISO Communication Under NR-RIS CRACK}
In a TDD implementation, the BS first estimates the CSI by processing uplink pilot signals transmitted by the $\rm{LUs}$. Leveraging the assumption of channel reciprocity, where the uplink and downlink channels are presumed to be identical, the BS then utilizes the estimated uplink CSI to design an appropriate precoder for downlink transmission (DT).

\subsubsection{Uplink Channel Estimation}Assuming the pilot signal transmitted by $\rm{LU_k}$ at time $t$ is denoted by $s_{t,k}$, the received signal from all of the $\rm{LUs}$ at the BS is given by
\begin{equation}\label{e1}
\mathbf{y}_{t}=\sum_{k=1}^K \sqrt{p_k}\mathbf{h}_{k,b}^{\star}s_{t,k}+\mathbf{n}_t,
\end{equation}
where $p_k$ is the power of the pilot signal assuming $E(|s_{t,k}|^2)=1$ and $\mathbf{n}_t$ denotes noise. In the presence of the NR-RIS, the uplink channel for $\rm{LU_k}$ is modeled as $\mathbf{h}^{\star}_{k,b}=\mathbf{H}_{r,b}(\mathbf{\Phi}-{\rm \mathbf{I}}_N )\mathbf{h}_{k,r}+\mathbf{h}_{k,b}$, which includes both the direct path $\mathbf{h}_{k,b}$ and the NR-RIS induced path\footnote{It has recently been shown via network-level multiport analysis that an additional ``structural scattering'' term is required to describe the full effect of RIS-induced signal propagation~\cite{10551389,li2024non}. Structural scattering corresponds to the virtual direct link that arises because the ``OFF'' state of the RIS is realized through residual current flow within the RIS circuitry, causing electromagnetic waves to scatter even when the RIS is inactive. This structural scattering term is overlooked in most RIS-aided communication studies, but it can impact practical RIS implementations, necessitating its consideration in system design. This is reflected in the cascaded channel model by an overall scattering matrix of $\mathbf{\Phi}-\mathbf{I}_N$ rather than simply $\mathbf{\Phi}$.}. Here, $\mathbf{I}_N$ denotes the $N\times N$ identity matrix, and $\mathbf{\Phi}$ is the scattering matrix of the NR-RIS, satisfying $\mathbf{\Phi} \mathbf{\Phi}^H = \mathbf{I}_N$ and $\mathbf{\Phi} \ne \mathbf{\Phi}^T$, indicating unitary but non-symmetric (i.e., non-reciprocal) behavior. We assume the channels $\mathbf{h}^{\star}_{k,b}$ for all users are known to the BS based on the uplink pilots, and the overall uplink channel is denoted as $\mathbf{H}_{up}=[\mathbf{h}^{\star}_{1,b},\cdots,\mathbf{h}^{\star}_{K,b}] = \mathbf{H}_{r,b}(\mathbf{\Phi}-{\rm \mathbf{I}}_N ) \mathbf{H}_{u,r}+\mathbf{H}_{u,b}\in \mathbb{C}^{M \times K}$, where $\mathbf{H}_{u,r}=[\mathbf{h}_{1,r},\cdots,\mathbf{h}_{K,r}]$ and $\mathbf{H}_{u,b}=[\mathbf{h}_{1,b},\cdots,\mathbf{h}_{K,b}]$ are the combined user-RIS channel and combined user-BS direct channel, respectively. 
 
\subsubsection{Downlink Transmission under CRACK}
Based on channel reciprocity, the BS assumes that the downlink channel is $\mathbf{H}_{down}=\mathbf{H}_{up}^T = \mathbf{H}_{u,r}^T(\mathbf{\Phi}-{\rm \mathbf{I}}_N )^{T} \mathbf{H}_{r,b}^T+\mathbf{H}_{u,b}^T$. However, due to the NR-RIS, the {\em actual} downlink channel is $\mathbf{H}_{down}^{\star}= \mathbf{H}_{u,r}^T(\mathbf{\Phi}-{\rm \mathbf{I}}_N ) \mathbf{H}_{r,b}^T+\mathbf{H}_{u,b}^T$. For a conventional RIS, $\mathbf{H}_{down}=\mathbf{H}^{\star}_{down}$ since the scattering matrix of a conventional RIS is diagonal, and hence symmetric. However, this is not the case for an NR-RIS since $\mathbf{\Phi} \neq\mathbf{\Phi}^T$, which breaks the channel reciprocity. The BS can only estimate the overall channel $\mathbf{H}_{up}$ and not its constituent parts $\mathbf{H}_{u,b}$ and $\mathbf{H}_{r,b}(\mathbf{\Phi}-{\rm \mathbf{I}}_N )\mathbf{H}_{u,r}$, which are not separately identifiable based only on uplink pilots. These terms can be estimated separately in a system with a cooperative RIS only because the RIS changes its response in a known way during training. The NR-RIS response is unknown to the BS and need not change with time to implement a covert attack. The inability to separately determine $\mathbf{H}_{r,b}(\mathbf{\Phi}-{\rm \mathbf{I}}_N ) \mathbf{H}_{u,r}$ and $\mathbf{H}_{u,b}$ prevents the BS from directly nulling the channel from the NR-RIS. Based on the assumed downlink channel $\mathbf{H}_{down}$, the BS designs the downlink precoder $\mathbf{W}=[\mathbf{w}_1, \cdots, \mathbf{w}_K]$ to provide data service during the DT phase, and the signal vector received by the $K$ users is given by 
\begin{equation}\label{e2}
\mathbf{y}=\mathbf{H}^{\star}_{down}\mathbf{W}\mathbf{s}+\mathbf{n}_d,
\end{equation}
where $\mathbf{n}_d$ denotes the noise received during DT, whose elements are modeled as independent and identically distributed (i.i.d.) random variables with zero mean and variance $\sigma^2$. 

\subsection{Achievable Rate under MRT and ZF Precoding}
To explore the impact of NR-RIS CRACK on TDD MU-MISO systems, two widely used linear precoding technologies, maximum ratio transmission (MRT) and zero-forcing (ZF) precoding are adopted in the simulations to facilitate performance comparisons with other related works~\cite{10081025,10149173,10424421,10445725}.
The respective precoding matrices for MRT and ZF are 
\begin{align}
\mathbf{W}_{\rm MRT} & =\xi \left(\mathbf{H}_{up}^T\right)^H=\xi \mathbf{H}^*_{up}=[\mathbf{w}_{\rm MRT,1},\dots,\mathbf{w}_{{\rm MRT},K}] \\
\mathbf{W}_{\rm ZF} &=\xi \mathbf{H}^*_{up}\left(\mathbf{H}^T_{up}\mathbf{H}^*_{up}\right)^{-1}=[\mathbf{w}_{\rm ZF,1},\dots,\mathbf{w}_{{\rm ZF},K}] \; ,
\end{align}
where $\xi$ is a scaling factor chosen to satisfy the transmit power constraint at the BS.

As explained above, the actual channel from the BS to $\rm{LU_k}$, which we denote as the $1\times M$ vector $\mathbf{h}^{\star}_{b,k}$, is given by $\mathbf{h}^{\star}_{b,k}=\mathbf{h}^T_{k,r}(\mathbf{\Phi}-{\rm \mathbf{I}}_N ) \mathbf{H}^T_{r,b}+\mathbf{h}^T_{k,b}$. Thus, the received downlink signal at $\rm{LU_k}$ is written as
\begin{equation}\label{e3}
y_k=\mathbf{h}^{\star}_{b,k} \mathbf{w}_{k}s_k+\mathbf{h}^{\star}_{b,k}\sum_{i=1, i\neq k}^{K}\mathbf{w}_{i}s_i + n_k.
\end{equation}
For notational simplicity, we write $\mathbf{h}_{r,k}=\mathbf{h}^T_{k,r}$, $\mathbf{H}_{b,r}=\mathbf{H}^T_{r,b}$, and $\mathbf{h}_{b,k}=\mathbf{h}^T_{k,b}$, so that the downlink signal-to interference-plus-noise ratio (SINR) for $\rm{LU_k}$ is given by 
\begin{equation}\label{e4}
\eta_k=\frac{|(\mathbf{h}_{r,k}(\mathbf{\Phi}-{\rm \mathbf{I}}_N ) \mathbf{H}_{b,r}+\mathbf{h}_{b,k})\mathbf{w}_k|^2}{\sum_{i=1, i\neq k}^{K}|(\mathbf{h}_{r,k}(\mathbf{\Phi}-{\rm \mathbf{I}}_N )\mathbf{H}_{b,r}+\mathbf{h}_{b,k})\mathbf{w}_i|^2+\sigma^2},
\end{equation}
and the achievable rate for $\rm{LU_k}$ can be written as 
\begin{equation}\label{e5}
r_k = \log(1+\eta_k).
\end{equation}
Due the NR-RIS, the BS uses an inaccurate downlink channel estimate, resulting in a mismatched precoding matrix and degraded performance. For example, under ZF precoding, the inter-user interference term $\sum_{i=1, i\neq k}^{K}|(\mathbf{h}_{r,k}(\mathbf{\Phi}-{\rm \mathbf{I}}_N 
)\mathbf{H}_{b,r}+\mathbf{h}_{b,k})\mathbf{w}_{{\rm ZF},i}|^2$ in \eqref{e4} is no longer zero because the orthogonality between $\mathbf{h}^{\star}_{b,k}$ and $\mathbf{w}_{{\rm ZF},i}$ is lost. 

\subsection{Physically-Consistent NR-RIS Structure}\label{sec:phycon}

\begin{figure}[!t]
\begin{center}
\includegraphics[width=3.5in]{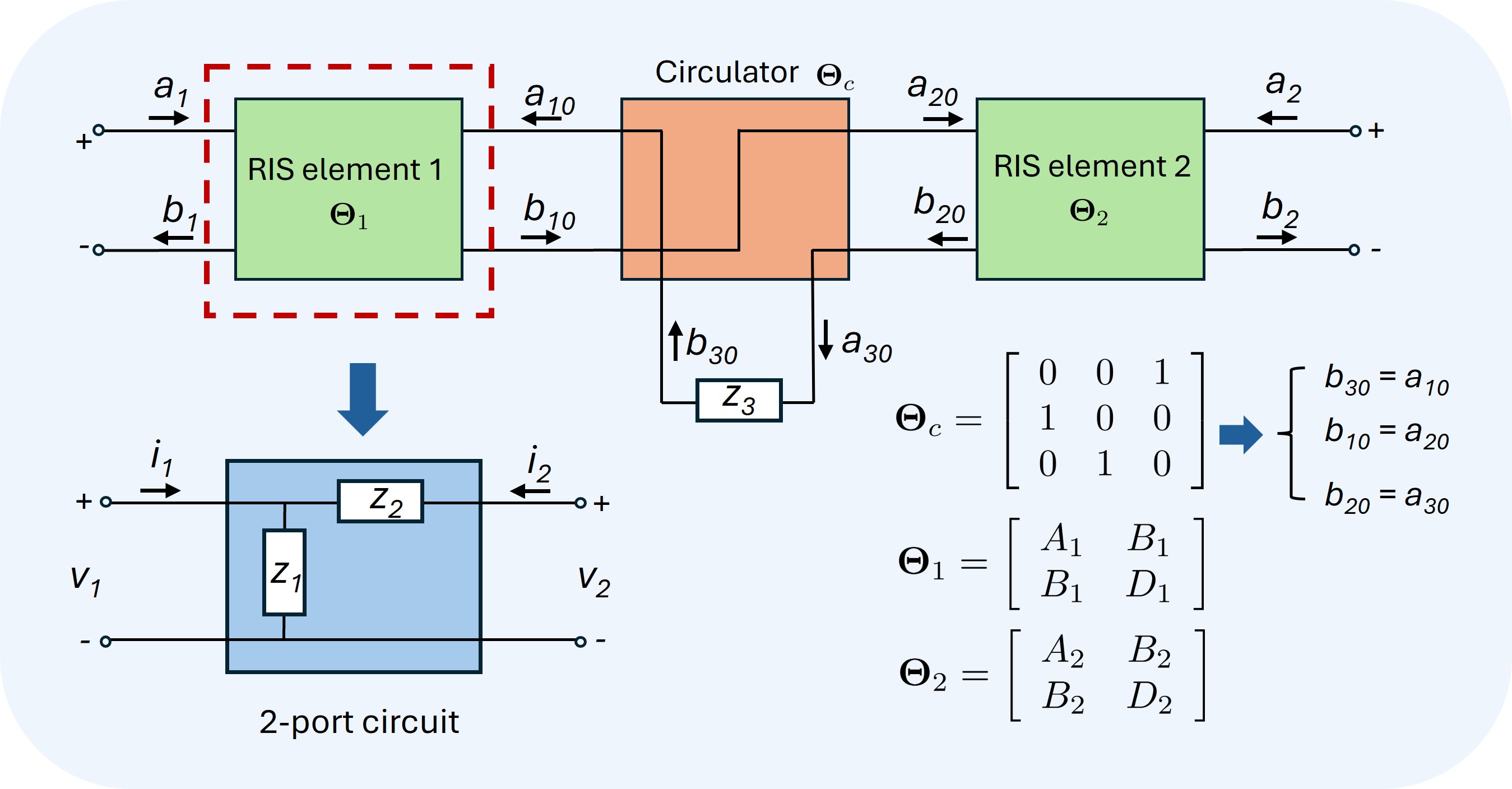}
\end{center}
\vspace{-0.08in}
\caption{Illustration of a NR dual-element unit.}\label{f2}
\vspace{-0.13in}
\end{figure}

A beyond-diagonal RIS (BD-RIS) leverages reconfigurable impedance networks with additional tunable elements to realize non-diagonal scattering matrices, offering greater wave control than conventional RIS~\cite{10453384}. Incorporating non-reciprocal devices (e.g., isolators, circulators) in a BD-RIS enables non-reciprocal scattering behavior~\cite{10694006,nrris}. In the discussion below, we introduce two simple physically-consistent NR-RIS models for CRACK. While other more general NR-RIS architectures are possible, they require a significant increase in implementational complexity, and these simple approaches are already sufficient to produce a significant impact.

\subsubsection{NR Dual-Element Unit} 
Assume a two-port model for each RIS element. Since the RIS element without any NR connections is reciprocal, the scattering matrix of $n$-th element is characterized by a $2\times 2$ reciprocal matrix:
\begin{equation}\label{e6}
    \mathbf{\Theta}_n = \begin{bmatrix}
A_n & B_n \\
B_n & D_n
\end{bmatrix} = (\mathbf{Z}_n-\mathbf{I}Z_0)(\mathbf{Z}_n+\mathbf{I}Z_0)^{-1} ,
\end{equation}
where $Z_0$ is the free-space impedance and $\mathbf{Z}_n$ is the $2\times 2$ reciprocal impedance matrix of element $n$. 
As shown in  Fig.~\ref{f2}, a 3-port circulator with scattering matrix denoted by $\mathbf{\Theta}_c$ that connects two RIS elements and a tunable reactive impedance $Z_3$ can achieve non-reciprocal characteristics. As shown in~\cite{nrris}, by properly choosing the RIS element impedances and the terminating impedance $Z_3$, the equivalent $2\times 2$ scattering matrix for the unit can be made to satisfy  
\begin{equation}\label{e7}
    \mathbf{\Phi} = \begin{bmatrix}
0 & e^{j\varphi_1} \\
e^{j\varphi_2} & 0
\end{bmatrix},
\end{equation}
where $\varphi_1, \varphi_2 \in (0, 2\pi]$ are arbitrary.
This basic structure is referred to as a NR dual-element unit. To strengthen the non-reciprocity, the heuristic setting $\varphi_1-\varphi_2=\pi$ can be adopted.

\subsubsection{NR-RIS Model with Block Structure}\label{nrris_structure}
In this architecture, we assume the NR-RIS is grouped into $G$ blocks of adjacent elements that can be controllably connected pairwise as NR dual-element units. The scattering matrix for the NR-RIS is thus constrained to have the form $\mathbf{\Phi}={\rm blkdiag}\left\{\mathbf{\Phi}_1,\cdots, \mathbf{\Phi}_g,\cdots, \mathbf{\Phi}_G\right\}$, where subblock $\mathbf{\Phi}_g$ is $L_g \times L_g$ and has $L_g/2$ NR dual-element units. The overall NR-RIS thus has $N=\sum_g L_g$ elements. For example, let $L_g=6$ and suppose that the element pairing strategy is given by $\left\{(1,5),(2,3),(4,6)\right\}$. Then
\[
\mathbf{\Phi}_g=\left[
\begin{array}{cccccc}
0 & 0 & 0 & 0 & e ^{j\varphi_{1,5}} & 0 \\
0 & 0 & e ^{j\varphi_{2,3}} & 0 & 0 & 0 \\
0 & e ^{j\varphi_{3,2}} & 0 & 0 & 0 & 0\\
0 & 0 & 0 & 0 & 0 & e ^{j\varphi_{4,6}}\\
e ^{j\varphi_{5,1}} & 0 & 0 & 0 & 0 & 0\\
0 & 0 & 0 &  e ^{j\varphi_{6,4}} & 0 &0\\
\end {array}
\right]
\]
with arbitrary phases. We assume the pairing and phase values for each block can be arbitrarily reconfigured. This motivates the desire to keep the size of the blocks relatively small to avoid the need for a prohibitively complex network of switches and interconnections. Our simulations in Section~\ref{sec:sim} will illustrate that the NR-RIS can be effective without requiring arbitrary reconfigurability across the entire surface.

\section{Covert NR-RIS CRACK}\label{s3}

Similar to conventional RIS-based attacks in \cite{10081025,10149173,10424421,10682037}, NR-RIS CRACK can be implemented entirely passively, without signal transmissions, and in general it does not require CSI from the legitimate system. However, CRACK offers several key advantages over these methods. The attacks in \cite{10081025,10149173,10424421,10682037} achieve the discrepancy $\mathbf{H}_{down} \ne \mathbf{H}_{up}^T$ by dynamically adjusting the phase shifts of a diagonal (and hence reciprocal) RIS during uplink training and downlink transmission, inducing ``active channel aging" (ACA) that reduces the channel coherence time and causes faster channel variations than assumed by the BS. Compared to CRACK, ACA attacks have two major limitations. First, ACA attackers must tightly synchronize with the uplink and downlink phases of the legitimate TDD system, which may be hard to realize in practice. Second, the induced rapid channel variations are relatively easy to detect and mitigate using polluted path separation techniques~\cite{10500386,10247269}. 

In contrast, NR-RIS CRACK represents a more covert strategy since the NR-RIS is naturally non-reciprocal and can be seamlessly integrated into the communication environment without actively introducing detectable interference or rapid phase changes. It does not affect the channel coherence time and remains physically undetectable, except through the resulting degradation in system performance. In what follows below, we outline three different types of NR-RIS CRACK threats and summarize their characteristics. The first two do not rely on any particular knowledge about the legitimate system, while the third presents a heuristic method for taking such information into account when available. We will see that, in addition to deteriorating the system throughput, CRACK can also increase the likelihood of security risks. 

\subsection{Blind CRACK} 
Blind CRACK refers to the standard scenario where the adversary lacks prior knowledge of the channel statistics for the legitimate system, and attempts to degrade the downlink throughput in a covert manner. In this case the NR-RIS employs a strategy that does not impact the existing coherence time of the channel, unlike some strategies based on diagonal RIS~ \cite{10149173,10424421,10682037} in which the reconfiguration interval $\triangle t$ of the RIS $\Delta t$ is significantly shorter than the channel coherence time $\Delta_c$. Conversely, the NR-RIS scattering matrix $\mathbf{\Phi}$ either remains static or randomly changes at a rate satisfying $\Delta t \simeq \Delta_c$ so as not to change $\Delta c$ and make the CRACK attack appear to be a naturally occurring environmental factor.

\subsection{CRACK-Aided Eavesdropping} 
Due to the broadcast nature of the physical medium, multi-user wireless communications are very susceptible to eavesdropping attacks. CRACK implemented in the presence of an eavesdropper can not only deteriorate the system rate, but also lead to information leakage from the legitimate system\footnote{Some traditional jamming-assisted eavesdropping strategies increase the probability of eavesdropping by transmitting jamming signals to reduce the target users' rate~\cite{7321779,9305288}. The strategy of using pilot contamination to enable active eavesdropping requires both pilot and jamming signal spoofing~\cite{6151778}. In~\cite{9576644}, a passive pilot spoofing attack is achieved by configuring the RIS with different values during uplink and downlink transmissions to deteriorate the secrecy rate. However, such approaches are easier to expose or require real-time synchronization with the system under attack. CRACK-aided eavesdropping is more covert due to its ability to be tuned at a rate consistent with the surrounding propagation environment.}. Since CRACK distorts the precoding design at the BS, it causes downlink signal energy to deviate from the intended legitimate users, which not only reduces the effectiveness of legitimate communication but also increases the risk of information leakage. Such an approach need not require knowledge of the eavesdropper’s channel. While better eavesdropping performance could be achieved if the eavesdropper’s channel were known, this would require the eavesdropper to be active rather than passive. 
To illustrate this case, in the simulations we will consider a common security threat scenario in which a single-antenna eavesdropper can monitor the transmissions to any of the legitimate users. The secrecy rate for ${\rm LU}_k$ is defined as $r_{s,k}=[r_k -r_{e,k}]^+$, where $[x]^+={\rm max}\left\{0,x\right\}$ and $r_{e,k}$ is the resulting rate achieved by the eavesdropper in intercepting ${\rm LU}_k$'s messages. A secrecy outage occurs if $r_{k} < r_{e,k}$, which means that information theoretic confidentiality cannot be guaranteed, and hence the eavesdropper may fully recover the messages destined by ${\rm LU}_k$. Even when the NR-RIS has no information about the legitimate system, CRACK will enhance the probability of eavesdropping success by severely lowering the legitimate users' rate.

\subsection{Knowledge-driven CRACK}
Many studies on communication security adopt a worst-case assumption: the adversary has perfect instantaneous CSI of the legitimate system to launch jamming or signal cancellation attacks~\cite{9112252,10302337,10402016}, which often leads to overly conservative remedies. Instead, in addition to the case where the adversary has no CSI for the legitimate system, we consider the case where the adversary can acquire at least partial CSI (e.g., user locations, path loss, or LoS components, etc.) via long-term observation of the network. Although exact rate expressions are in general difficult to derive and optimize, the CSI may enable the adversary to heuristically identify a set of effective NR-RIS configurations ${\mathbf{\Phi}^{\dagger}}$ to strengthen the attack performance. In the simulations, we will present a simple but effective method that configures the NR-RIS to maximize the difference between the uplink and downlink BS-RIS-User LoS channel components.

\subsection{Countermeasure Problem Formulation and Challenges}
We will see in the simulations that the CRACK attack strategies described above are very effective in degrading the performance of the legitimate systems when they employ conventional precoding. Thus, if the BS determines that an NR-RIS implementing CRACK is present during coherence interval $t$, it should attempt to design an alternative precoder $\mathbf{W}_t$ that is more robust. For example, such a precoder could attempt to maximize the sum rate subject to transmit power constraints, as in the following optimization:
\begin{equation}
\max _{\mathbf {W}_t} \sum_{k=1}^{K} r_{t,k} \quad \text{s.t.}\quad \sum_{k=1}^{K} \|\mathbf{w}_{t,k}\|^2= P_{\rm total},
\end{equation}
where $r_{t,k}$ is the achievable rate for user $k$ in coherence block $t$, and $P_{\rm total}$ is the total transmit power at the BS. Since the NR-RIS passively and gradually alters the channel environment, this is a challenging problem to solve via conventional TDD-based methods. To mitigate the impact of NR attacks on legitimate system performance and security, we propose the use of deep reinforcement learning (DRL) techniques, which if properly trained can handle NR attacks in highly dynamic scenarios, especially with imperfect CSI. In the next section we design a DRL-based network that is trained using rate feedback from the users to learn a more robust precoder in the presence of CRACK.

\section{DRL-Based Countermeasures}
In this section, we propose a DRL-based approach to address the problem of robust precoder design to address the possible effect of a malicious NR-RIS. We first analyze the challenges associated with using DRL and propose an enhanced proximal policy optimization (PPO) algorithm with several improvements. Then we introduce our SecureCoder algorithm and describe its implementation in detail\footnote{The source code for our proposed SecureCoder with the test environment will be available soon.} 

\subsection{DRL: Challenges and Solutions}
DRL offers a promising solution for the problem under consideration since it operates by interacting with the environment and learning complex, non-intuitive mappings. However, its application in this scenario faces challenges. First, the dynamic wireless environment and the unpredictable impact of an NR-RIS require rapid adaptation to time-varying channel effects. Second, the BS only observes the uplink CSI while the NR-RIS distorts the downlink CSI, making the link between the observations and system performance implicit and difficult to model. As a result, reward signals (e.g., user rate) are often sparse and noisy, reducing training stability and increasing sensitivity to initialization. Third, both the input (uplink CSI) and output (precoding matrix) are continuous and high-dimensional, requiring an efficient DRL framework with strong generalization.
To restore a proper downlink precoder under CRACK, we propose the ``SecureCoder'' scheme, which is based on a tailored DRL framework, referred to as ``enhanced-PPO'', with the following key components:
\begin{enumerate}
\item Stable Policy Learning via PPO: We adopt the Proximal Policy Optimization (PPO) algorithm due to its robustness in high-dimensional and continuous action spaces~\cite{schulman2017ppo}. The clipped objective function in PPO helps constrain the policy update within a trust region, which is particularly beneficial under sparse or noisy reward conditions, preventing instability in training.
\item State Representation Enhancement via CNN: To process the high-dimensional uplink channel matrix used as the state input, we employ a convolutional neural network (CNN) as a front-end feature extractor. The CNN captures potential spatial correlations in the channel matrix, enabling the PPO to use more informative features. This CNN+PPO hybrid enhances both learning efficiency and generalization across diverse channel conditions.
\item Reward Sparsity Mitigation via Specialized PER:
We integrate a prioritized experience replay (PER) scheme into the PPO framework to accelerate policy learning. Since PPO is on-policy, incorporating PER requires careful handling. To address this, we selectively store high-reward trajectories from recent episodes and reuse them in future policy updates, while preserving advantage estimates based on their original rollouts. This approach combines the stability of the PPO’s clipped updates with the efficiency gains from reusing valuable past experiences.
\end{enumerate}

\subsection{DRL-Based SecureCoder}

\begin{figure*}[!t]
\begin{center}
\includegraphics[width=5.8 in]{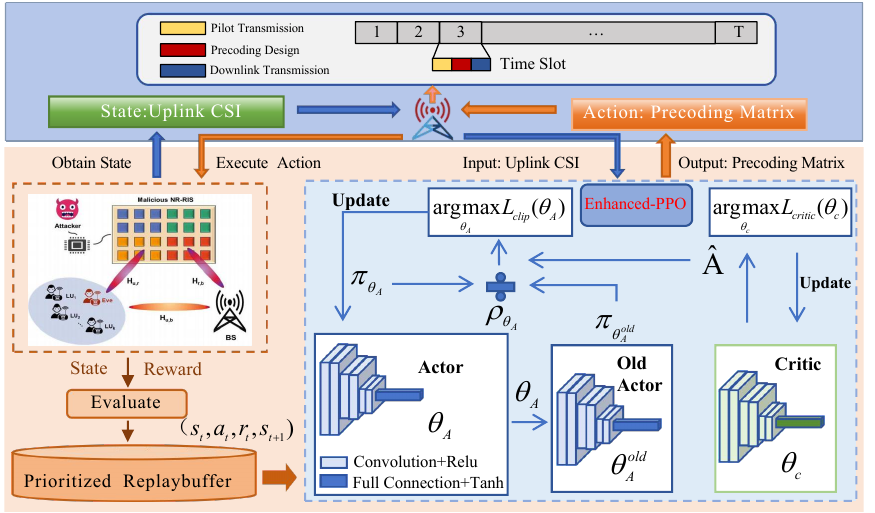}
\end{center}
\caption{Framework of DRL-based SecureCoder}\label{f3}  
\vspace{-0.13in}
\end{figure*}

Figure~\ref{f3} shows the DRL-based SecureCoder framework, where the BS acts as an intelligent agent that aims to learn a defensive transmission strategy against NR-RIS CRACK through iterative interaction with the wireless environment. At each channel coherence time $t$, the BS first estimates the uplink channel $\mathbf{H}_{up}$ which forms the state information $\mathbf{s}_t$ for the DRL structure. Accordingly, the agent designs the downlink precoding matrix denoted as action $\mathbf{a}_t$ and conducts communication with the legitimate users. Afterwards, the BS agent obtains reward feedback $r_t$ (defined below) from the users measuring the communication quality, after which the agent updates the overall system towards a new state $\mathbf{s}_{t+1}$ affected by the communication environment. These sequential interactions generate a series of experience trajectories $\left(\mathbf{s}_t, \mathbf{a}_t, r_t, \mathbf{s}_{t+1}\right)$ which are stored and used to update the action policy via specific DRL algorithms. The BS agent will balance the exploration of new actions with the exploitation of learned strategies to maximize the long-term cumulative reward. This continuous interaction with the sequential decision-making process can be systematically characterized by a Markov Decision Process (MDP). 
In the following subsections, we elaborate on the design of the core MDP components.

\subsubsection{State Space}
When the TDD system is subject to covert CRACK, the BS can only use the estimated uplink CSI for the precoding design. Assuming that the BS can perform precise channel estimation based on the received pilot signals, the state at interval $t$ is defined as $\mathbf{s}_t=\left\{\mathbf{H}^{ a}_{up,t}, \mathbf{H}^{ p}_{up,t}\right\}$, where the complex-valued uplink channel matrix $\mathbf{H}_{up,t}$ is separated into two real-valued parts, i.e., a matrix of amplitudes $\mathbf{H}^{ a}_{up,t}$ and matrix of phases $\mathbf{H}^{ p}_{up,t}$, both essential for representing the channel state.
Compared to separating $\mathbf{H}_{up,t}$ into real and imaginary parts, splitting it into amplitude and phase preserves the physical interpretability of the complex-valued channel states while circumventing the numerical instability of directly processing complex tensors. By interacting with the environment, the CNN learns to extract the effective channel features from the CRACK-induced anomalies. Thus the CNN output provides a basis for the BS agent to formulate appropriate precoding strategies.

\subsubsection{Action Space}
Once the BS agent obtains the state information, it determines the action $\mathbf{a}_t$ at time $t$, i.e., the downlink precoding matrix, defined as $\mathbf{a}_t=\left\{\mathbf{W}_t \right\}$. 
To deal with the complex-value constraints on $\mathbf{W}_t$, the network generates two real-valued matrices $\mathbf{W}_t^{a}$ and 
$\mathbf{W}_t^{p}$ representing the amplitude and phase components of the precoder, respectively. For the DRL process, the elements of these matrices are constrained to lie in the range $[0, 1]$ by modeling the output distributions of the actor network using Beta distributions. Before generating the precoder, $\mathbf{W}_t^{a}$ is normalized as $\sqrt{P_{\text{total}}} \mathbf{W}_t^{a} / \|\mathbf{W}_t^{a}\|_{\rm F}$ to satisfy the total transmit power constraint. Then, the complex-valued precoding matrix $\mathbf{W}_t$ is realized as $\mathbf{W}_t=\mathbf{W}_t^{a}\odot e^{j\cdot 2\pi \mathbf{W}_t^{p}}$, where $\odot$ represents an element-wise product. As with the state information, this representation aligns well with the physical characteristics of wireless signals allowing the network to learn the signal power and directionality independently, which enhances interpretability and control. It also facilitates more stable training, as the amplitude and phase values are non-negative and lie within a bounded range, improving the learning efficiency of the PPO agent.

\subsubsection{Reward Function}
An effective reward function is crucial for guiding the DRL agent towards mitigating the impact of NR-RIS CRACK. While sum rate is an obvious choice, it often leads to unequal resource allocation, favoring users with strong channels and marginalizing others. To avoid this, we define the reward function using a logarithm applied to each user's achievable rate before summation: 
\begin{equation}\label{e12}
r_t=\sum_{k = 1}^{K}\log(1 + r_{t,k}).
\end{equation}
This reward function guides the DRL agent to learn an effective precoding strategy that enhances the sum rate while promoting equitable resource allocation. Increasing the SINR of users with poor channel conditions indirectly reduces the likelihood of successful eavesdropping, so the system is able to achieve a balance between throughput efficiency, fairness, and secrecy outage under adversarial conditions.

\subsection{Enhanced-PPO Algorithm}
The enhanced PPO framework extends the PPO algorithm by integrating a CNN into the actor-critic architecture. While conventional PPO directly maps input states to actions using fully connected layers, enhanced PPO introduces a CNN-based front end to both the actor and critic networks. Specifically, the complex-valued CSI matrix is first transformed into a real-valued tensor and passed through two convolutional layers to extract the effective spatial and structural features of the channel.
These learned features are subsequently fed into fully connected layers, where the actor network generates a stochastic policy for precoding decisions, and the critic network estimates the state value function for policy evaluation. This architectural enhancement allows the actor to better capture the underlying characteristics of dynamic wireless channels, thereby facilitating more efficient and robust policy learning. 

Beyond network design, the enhanced PPO adheres to the same principle as standard PPO. The PPO algorithm imposes a limit on the probabilities of the new and previous policies, ensuring that policy updates lie within a bounded range. To this end, PPO uses the following surrogate objective
that constrains policy updates:
\begin{equation}\label{e13}
\begin{aligned}
J^{\mathrm{clip+ent}}\left(\boldsymbol{\theta}_{\mathrm{A}}\right) = \hat{\mathbb{E}}_t\left[\min \left(\rho_t \hat{A}_{t}, \text{clip}(\rho_t, 1-\varepsilon, 1+\varepsilon) \hat{A}_{t}\right)\right] \\
+ \beta H\left(\pi_{\boldsymbol{\theta}_{\mathrm{A}}}\right),
\end{aligned}
\end{equation}
where $\boldsymbol{\theta}_{\mathrm{A}}$ is the vector of actor network parameters and $\rho_t = \frac{\pi_{\boldsymbol{\theta}_{\mathrm{A}}}(\mathbf{a}_{t} \mid \mathbf{s}_{t})}{\pi_{\boldsymbol{\theta}_{\mathrm{A}}^{\text {old }}}(\mathbf{a}_{t} \mid \mathbf{s}_{t})}$ is the importance sampling ratio between the new policy $\pi_{\boldsymbol{\theta}_{\mathrm{A}}}$ and the old policy $\pi_{\boldsymbol{\theta}_{\mathrm{A}}^{\text {old }}}$ used to collect data trajectories. The term $\hat{A}_{t}$ is the advantage function, which evaluates the action quality of a specific state. The $\text{clip}(\cdot)$ function ensures that the updates to the policy stay within the range $[1-\varepsilon, 1+\varepsilon]$, effectively mitigating drastic changes that could lead to unstable behavior or performance degradation.
To encourage policy exploration and prevent premature convergence, a term $H\left(\pi_{\boldsymbol{\theta}_{\mathrm{A}}}\right)$ is introduced that represents the expected entropy of the action distribution: 
\begin{equation}\label{e14}
H\left(\pi_{\boldsymbol{\theta}_{\mathrm{A}}}\right) = \hat{\mathbb{E}}_t \left[ -\pi_{\boldsymbol{\theta}_{\mathrm{A}}}\left(\mathbf{a}_{t} \mid \mathbf{s}_{t}\right) \log \pi_{\boldsymbol{\theta}_{\mathrm{A}}}\left(\mathbf{a}_{t} \mid \mathbf{s}_{t}\right) \right].
\end{equation}
By adjusting the entropy coefficient $\beta$, the policy update achieves a balance between exploration and exploitation.

During the training phase, the agent leverages recently stored trajectories from the experience replay buffer to update the parameters of the actor network $\theta_{\mathrm{A}}$ that maximizes the surrogate objective $J^{\mathrm{clip+ent}}(\theta_{\mathrm{A}})$ while maintaining policy stability within a trust region:  
\begin{equation}\label{e15}
\begin{aligned}
\boldsymbol{\theta}_{\mathrm{A}}^{\text{new}} = \arg\max_{\boldsymbol{\theta}_{\mathrm{A}}} \, J^{\mathrm{clip+ent}}\left(\boldsymbol{\theta}_{\mathrm{A}}\right).
\end{aligned}
\end{equation}
The critic network is optimized to accurately estimate the state by leveraging temporal difference (TD) learning, which computes the target value $V^{\text{target}}(\mathbf{s}_t)$ for each state $\mathbf{s}_t$ as
\begin{equation}\label{e16}
V^{\text{target}}(\mathbf{s}_t) = r(\mathbf{s}_t, \mathbf{a}_t) + \gamma V_{\theta_{\mathrm{C}}}(\mathbf{s}_{t+1}),
\end{equation}
where $\gamma$ is the discount factor. 
The parameters of the critic network $\theta_{\mathrm{C}}$ are then updated using a mean squared error loss function that minimizes the discrepancy between the predicted value $V_{\theta_{\mathrm{C}}}(\mathbf{s}_t)$ and the TD-derived target:
\begin{equation}\label{e17}
 \theta_{\mathrm{C}}^{\text{new}} = \theta_{\mathrm{C}} - \alpha_{\mathrm{c}} \cdot \frac{1}{I} \sum_{i=1}^I \nabla_{\theta_{\mathrm{C}}}\left(V_{\theta_{\mathrm{C}}}(\mathbf{s}_t) - V^{\text{target}}(\mathbf{s}_t)\right)^2,
\end{equation}
where $\alpha_{\mathrm{c}}$ is the learning rate and $I$ is the number of trajectories in each batch. The experience replay of standard PPO uses a uniform random sampling mechanism. However, sparse reward feedback will reduce learning efficiency and stability. In contrast, the proposed enhanced PPO adopts a specialized prioritized experience replying scheme. For each parameter updating period, the agent selects a fixed proportion of high-reward trajectories from the recently updated experience reply buffer, while the rest continue to follow the random sampling mechanism, thereby enhancing sample efficiency. 

\addtolength{\topmargin}{-0.01in}
 \begin{algorithm}[!t]
    \caption{Enhanced-PPO Framework }
    \label{alg:1}
    \begin{algorithmic}[1]
        \STATE \textbf{Initialization}: Orthogonally initialize the parameters of the actor network  $\theta_{\mathrm{A}}$ and the critic network $\theta_{\mathrm{C}}$. Initialize the prioritized experience buffer.
        \FOR{each episode $i=1,2,...,Z$}
            \STATE $\pi_{\theta_{\text {old }}} \leftarrow \pi_\theta$
            \FOR{each time slot $t=1,2,...,T$}
                \STATE Agent obtains its observations and executes action according to policy $\pi_{\theta_{\text {old }}}$. 
                \STATE Agent obtains the reward $r_{t}$ and proceeds to next state $\mathrm{s}_{t+1}$.
                \STATE Store trajectory $\left\{\mathbf{s}_{t}, \mathbf{a}_{t}, r_t, \mathbf{s}_{t+1}\right\}$ in the prioritized experience buffer.
                \STATE Evaluate experience priority.
            \ENDFOR
            \STATE Sample trajectories from the agent's prioritized experience buffer.
            \STATE Compute the advantage function $\hat{A}_{t}$.
            \STATE Compute the target state value $V^{\text{target}}$.
            \STATE Update $\theta_{\mathrm{A}}$ using \eqref{e15} and update $\theta_{\mathrm{C}}$ using \eqref{e17}.
        \ENDFOR
    \end{algorithmic}
\end{algorithm}

The overall proposed approach is outlined in Algorithm 1, describing the training procedure and the sequential execution of the operations for DRL optimization. In the next section, we study the impact of various CRACK scenarios on the performance of the legitimate system, and validate the benefit of the DRL-based algorithm in improving system performance.

\section{Numerical Results}\label{sec:sim}
In this section, we provide numerical results to evaluate the communication threats posed by NR-RIS CRACK. Then, we examine the performance of SecureCoder under CRACK. 

\subsection{Scenario and Channel Models}
We consider an MU-MISO system where the BS and the NR-RIS are each equipped with uniform linear arrays with half-wavelength element spacing. The BS is located at the 3D coordinates (5m, 35m, 20m), and the LUs are randomly distributed in a circular region centered at (5m, 0m, 2m) with 10m radius. An eavesdropper (Eve) is located at (6m, 5m, 2m), and the NR-RIS is deployed at (0m, 30m, 15m). One of the benchmark algorithms employs a jammer instead of an NR-RIS, and for this case the jammer is located at the same position as the NR-RIS, transmitting with power $P_{\rm J}=30$ dBm. The transmit power at the BS is also $P_{\rm total}=30$ dBm.

We assume a Rician fading model for all links. In particular, the channel between $\rm{LU_k}$ and the BS is expressed as
\begin{equation}\label{e8}
\mathbf{h}_{k,b}=\sqrt{\alpha_{k,b}}\left(\sqrt{\frac{\kappa_{k,b}}{1+\kappa_{k,b}}} \overline{\mathbf{h}}_{k,b}+\sqrt{\frac{1}{1+\kappa_{k,b}}} \widetilde{\mathbf{h}}_{k,b}\right) ,
\end{equation}
where $\alpha_{k,b}=\rho d_{k,b}^{-\iota_{k,b}}$ is the distance-dependent large-scale path-loss factor, $\rho$ is the path loss at a given reference distance, $d_{k,b}$ is the distance between $\rm{LU_k}$ and the BS, and $\iota_{k,b}$ is the path loss exponent of the $\rm{LU_k}$-to-BS link. The LoS channel component is denoted by $\overline{\mathbf{h}}_{k,b}$ which is assumed to satisfy $\overline{\mathbf{h}}_{k,b}^H\overline{\mathbf{h}}_{k,b}=M$, and the elements of the non-LoS (NLoS) component $\widetilde{\mathbf{h}}_{k,b}$ follow an i.i.d. complex Gaussian distribution with zero mean and unit variance. Similarly, the channel $\mathbf{h}_{k,r}$ between user $k$ and the NR-RIS is described as
\begin{equation}\label{e9}
\mathbf{h}_{k,r}=\sqrt{\alpha_{k,r}}\left(\sqrt{\frac{\kappa_{k,r}}{1+\kappa_{k,r}}} \overline{\mathbf{h}}_{k,r}+\sqrt{\frac{1}{1+\kappa_{k,r}}} \widetilde{\mathbf{h}}_{k,r}\right) , 
\end{equation}
where $\overline{\mathbf{h}}_{k,r}$ is the LoS channel satisfying $\overline{\mathbf{h}}_{k,r}^H\overline{\mathbf{h}}_{k,r}=N$. The eavesdropping-related channels $\mathbf{h}_{e,b}$ and $\mathbf{h}_{e,r}$ can be defined similarly according to (\ref{e8}) and (\ref{e9}), respectively. Finally, the BS-RIS channel $\mathbf{H}_{r,b}$ is described by 
\begin{equation}\label{e10}
\mathbf{H}_{r,b}=\sqrt{\alpha_{r,b}}\left(\sqrt{\frac{\kappa_{r,b}}{1+\kappa_{r,b}}} \overline{\mathbf{H}}_{r,b}+\sqrt{\frac{1}
{1+\kappa_{r,b}}} \widetilde{\mathbf{H}}_{r,b}\right),
\end{equation}
with LoS component $\overline{\mathbf{H}}_{r,b}=\sqrt{\alpha_{r,b}}\mathbf{a}_N \mathbf{a}_M^H$, where $\mathbf{a}_N^H\mathbf{a}_N=N$ and $\mathbf{a}_M^H\mathbf{a}_M=M$ represent the corresponding array response vectors. Unless otherwise specified, the path loss exponent of the RIS-BS link is $\iota_{r,b}=2$. 

We assume a physically consistent NR-RIS model that employs the simple NR dual-element unit architecture described in Section~\ref{nrris_structure}, where the random phases $\phi_1$ and $\phi_2$ satisfy $\phi_1-\phi_2=\pi$. Unless otherwise stated, we assume the NR-RIS allows arbitrary paired connections for which $G=1$ and $L=N$ in order to fully explore the system's performance. When using small-sized groups where $L_g < N$ for comparison, we will assume that all groups within the NR-RIS are of the same size, so that $L_g=L=N/G$.  Several different cases for the number of NR-RIS groups $G$ and BS antennas $M$ are considered for the simulations, as listed in Table~\ref{tab: SIMULATION PARAMETERS} together with other simulation parameters.

\begin{table}[!t]
    \vspace{-0.01in}
    \centering
    \caption{SIMULATION PARAMETERS}
    \label{tab: SIMULATION PARAMETERS}
    \begin{tabular}{|l|l|c|}\hline
        \textbf{Parameter}& \textbf{Value}\\\hline
        Number of users $K$ & 4 \\\hline
        Number of BS antennas $M$ & [8,16,32,64,128,256] \\\hline
        Number of RIS elements $N$ & [8,16,32,64,128,256] \\\hline
        Block size of NR-RIS elements $L$ & [2,4,8,16,32,64,128] \\\hline
       
        Path loss $\rho$ & $-20$ dB \\\hline
        Bandwidth $BW$ & $1$ MHz \\\hline
        Noise power $\sigma^2$ & $ 10^{-12} $~W \\\hline
        Path loss exponent $\iota_{k,r}$, $\iota_{k,b}$, $\iota_{e,r}$, $\iota_{e,b}$ & 2.5, 3.5, 2.5, 3.2
        \\\hline
        Rician factor $\kappa_{k,r}$, $\kappa_{k,b}$, $\kappa_{e,r}$, $\kappa_{e,b}$, $\kappa_{r,b}$ & 6, 3, 8, 4, 12 
        \\\hline
        Learning rate of actor and critic network $\alpha$& 0.005 \\\hline
        Entropy coefficient $\beta$ & 0.0001 \\\hline
        Discount factor $\gamma$ & 0 \\\hline
        Clipped parameter $\epsilon$ & 0.2 \\\hline
        Batch size $I$ & 2000 \\\hline
        Number of step $T$ & 20 \\\hline
        
    \end{tabular}
    \vspace{-0.01in}
\end{table}

\subsection{Benchmark Schemes and Metrics}
We compare the performance of a MU-MISO TDD system for the following scenarios: 
\begin{itemize}
\item[*] \textbf {No CRACK}: No RIS of any type is present.
\item[*] \textbf {NR-RIS}: An NR-RIS based on the physically consistent model described in Section~\ref{sec:phycon} is present, and a different random configuration is chosen for each coherence interval ($\Delta_t=\Delta_c$) for Blind CRACK and at the rate $\Delta_t=5\Delta_c$ for Knowledge-Aided CRACK.
\item[*] \textbf {Heuristic Algorithm (HA)}: Assumes LoS channels of the MU-MISO system are known. In each transmission period, the physically consistent NR-RIS configuration is chosen from among 200 random realizations to maximize the difference between the uplink and downlink LoS channels $\sum_{k=1}^{K}\beta_k\Vert \overline{\mathbf{H}}_{r,b}(\mathbf{\Phi}-{\rm \mathbf{I}}_N ) \overline{\mathbf{h}}_{k,r}-\overline{\mathbf{H}}_{r,b}(\mathbf{\Phi}^T-{\rm \mathbf{I}}_N ) \overline{\mathbf{h}}_{k,r}\Vert^2$, where $\beta_k=\frac{\alpha_{k,r}\alpha_{r,b}\kappa_{k,r}\kappa_{r,b}}{(1+\kappa_{k,r})(1+\kappa_{r,b})}$.
\item[*] \textbf {ND-RIS}: An idealized ND-RIS as assumed in~\cite{10445725} is present and a different random configuration is chosen for each coherence interval.
\item[*] \textbf {Diagonal RIS 1 (D-RIS 1)}: A conventional D-RIS is present and for each coherence interval is set to a random value during the PT phase, and then a different random value during the DT phase~\cite{10081025}.
\item[*] \textbf {D-RIS 2}: A conventional D-RIS is present that for each coherence interval is not active during PT and is set to a random value during DT~\cite{10149173}.
\item[*] \textbf {D-RIS 3}: A conventional D-RIS is present that for each coherence interval is set to a random value during the PT phase and then to multiple different random values during the DT phase~\cite{10424421}.
\item[*] \textbf {Jamming-Aided Eavesdropping (J-Eve)}: A single-antenna jammer is present that constantly emits interference with power $P_{\rm J}$ during the DT phase of each coherence interval. The eavesdropper is assumed to be able to eliminate the jamming interference.
\end{itemize}
In the following simulations, we generate 3000 random channel realizations to calculate the ergodic sum rate, ergodic sum secrecy rate, and secrecy outage probability (SOP). 

\subsection{Blind CRACK}

\begin{figure*}[!tp]
\begin{center}
\vspace{-0.05in}
\hspace*{-0.5cm}\includegraphics[width=6 in]{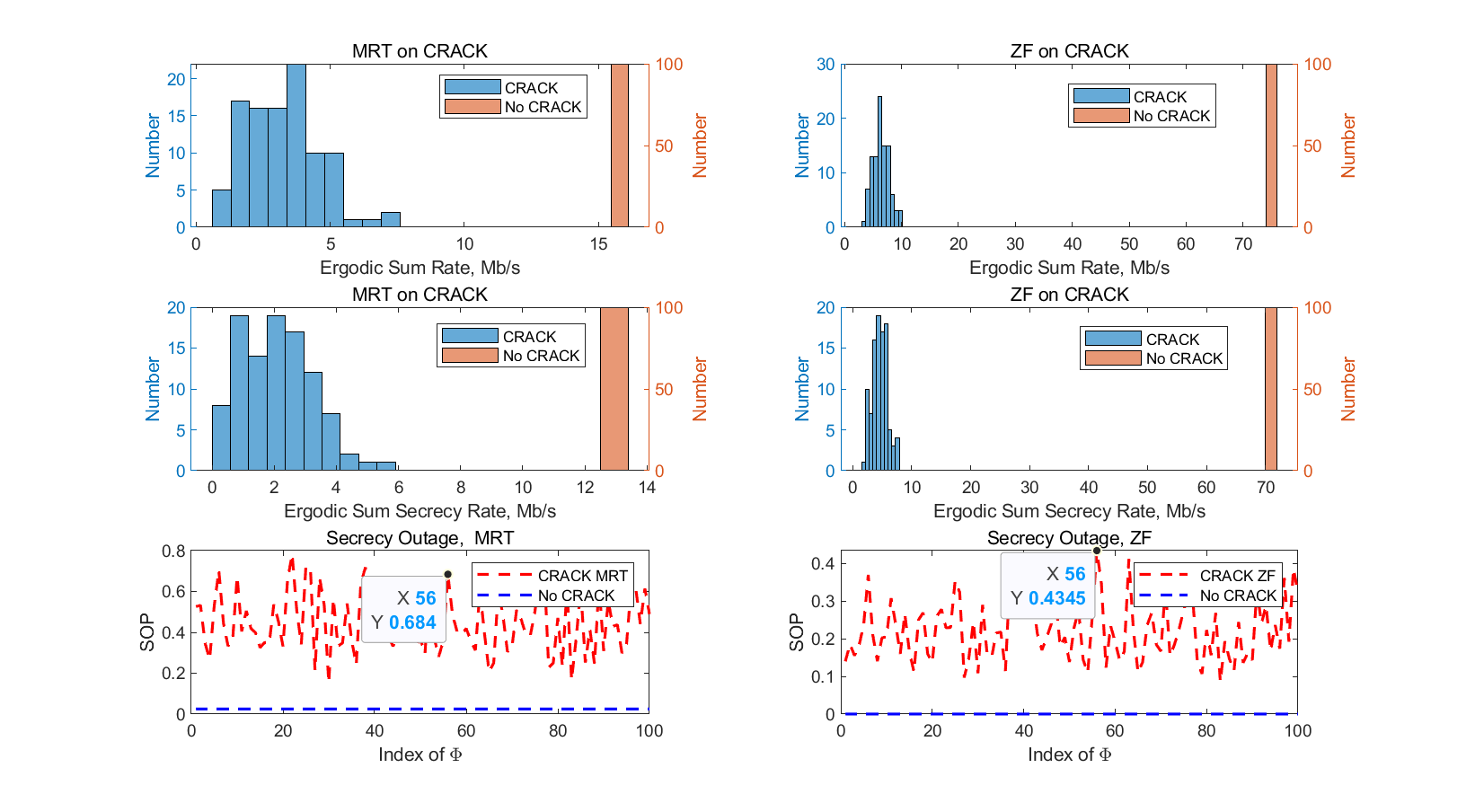}
\end{center}
\vspace{-0.2in}
\caption{Ergodic sum rate, sum secrecy rate, and SOP with and without NR-RIS for MRT and ZF, where N=128 and M=32.}\label{p1}
\vspace{0.05in}
\end{figure*}

\subsubsection{CRACK Performance}
In this example we generate 100 random NR-RIS configurations assuming $N=128$ and $M=32$, and for each configuration we calculate the ergodic sum rate, ergodic sum secrecy rate, and corresponding SOP of the MU-MISO system averaged over 3000 random channel realizations. As the resulting histograms in Fig.~\ref{p1} illustrate, while there are variations in the attack performance due to the random NR-RIS configurations, CRACK is very effective in negatively impacting the sum rate and sum secrecy rate of the legitimate network, particularly for ZF precoding. 
\begin{figure}[th]
\begin{center}
\vspace{-0.1in}
\includegraphics[width=3 in]{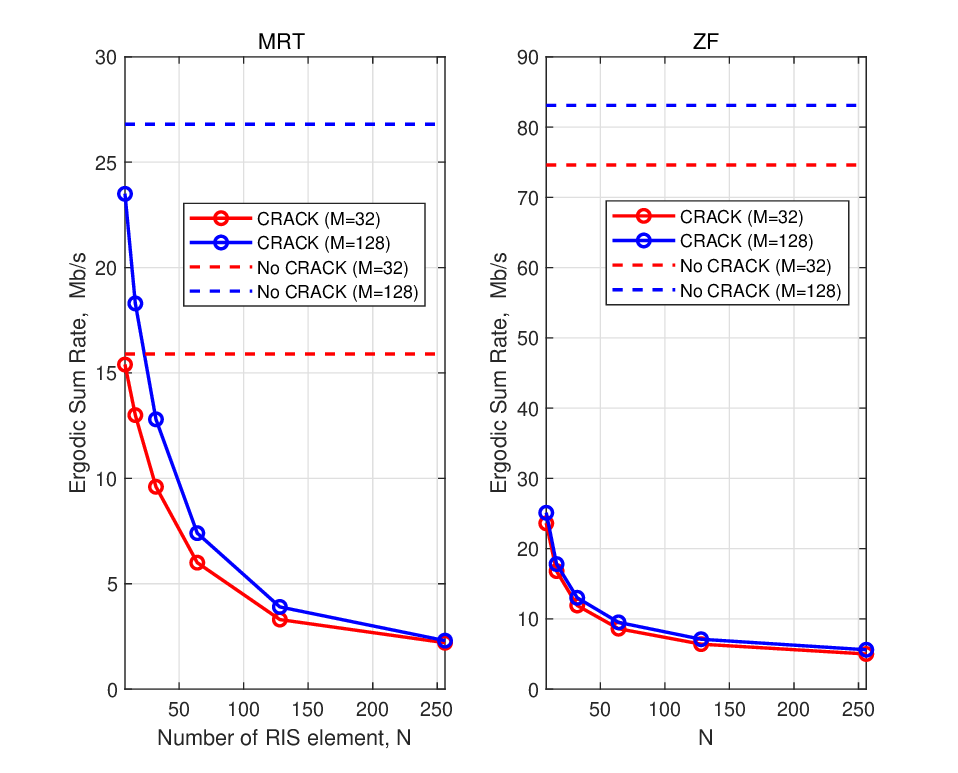}
\end{center}
\vspace{-0.1in}
\caption{Impact of BS array size on ergodic sum rate versus number of NR-RIS elements $N$.}\label{p2}
\vspace{-0.01in}
\end{figure}

\subsubsection{CRACK versus NR-RIS Size}
Fig.~\ref{p2} shows the ergodic sum rate versus the number of NR-RIS elements $N$ for $M = 32$ and $M = 128$ BS antennas. As $N$ increases, the sum rate for both MRT and ZF precoding drops sharply and eventually stabilizes at a very low level, even for large $M$. This demonstrates the severe performance degradation achieved by CRACK even without any CSI. For example, when $N = 256$ and $M = 128$, the ergodic sum rate drops by approximately $90\%$ under MRT and $92\%$ under ZF.
\begin{figure}[ht]
\begin{center}
\vspace{-0.1in}
\includegraphics[width=3 in]{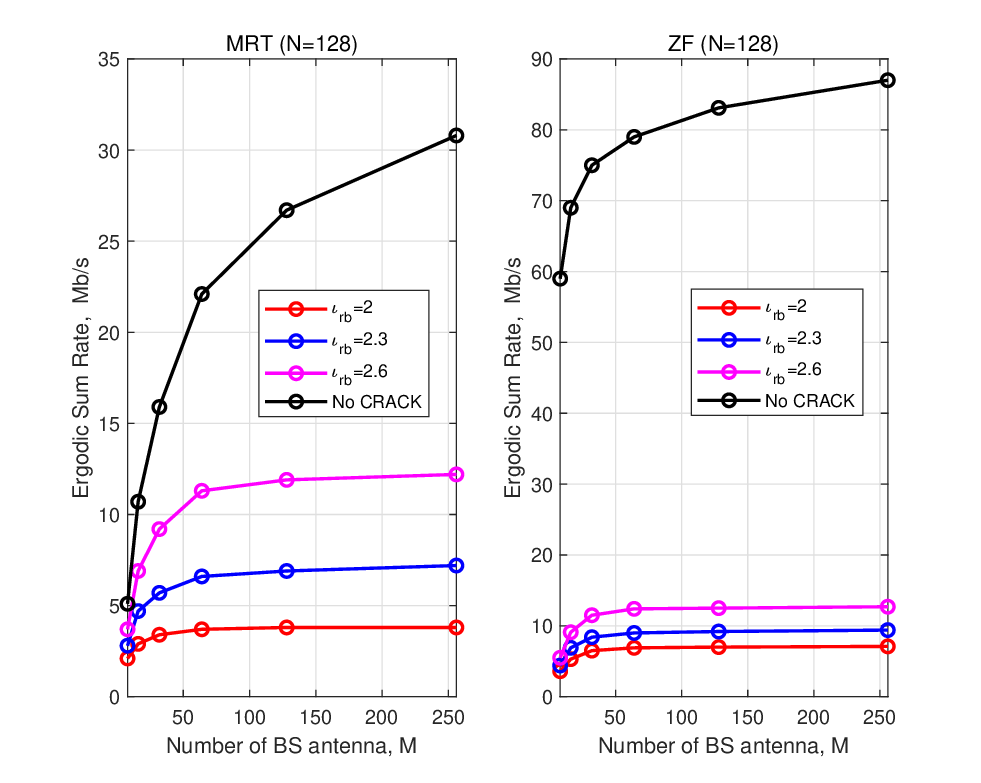}
\end{center}
\vspace{-0.1in}
\caption{Impact of BS-RIS path loss exponent $\iota_{r,b}$ 
on ergodic sum rate versus number of BS antennas $M$.}\label{p3}
\vspace{-0.11in}
\end{figure}

\subsubsection{CRACK versus BS Antenna Array Size}
Fig.~\ref{p3} plots the ergodic sum rate as a function of the number of BS antennas and highlights the influence of the BS-RIS channel path loss on CRACK performance. We see that beyond a certain point, increasing the size of the BS array cannot reduce the impact of CRACK. We also see that the effectiveness of the attack relies on the quality of the RIS-BS channel; higher path losses result in a diminished cascaded RIS channel gain and hence a weaker CRACK impact. Compared with MRT, ZF is more vulnerable as its reliance on channel orthogonality makes it highly sensitive to mismatched CSI, resulting in severe inter-user interference.

\begin{figure}[!t]
\begin{center}
\vspace{-0.01in}
\includegraphics[width=3 in]{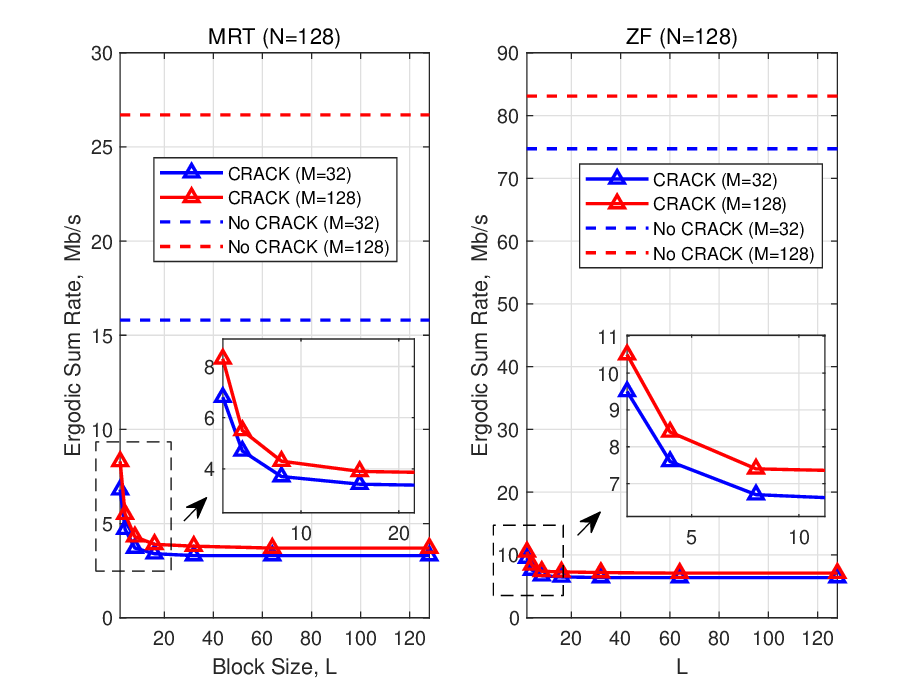}
\end{center}
\caption{Ergodic sum rate versus NR-RIS block size $L$. }\label{p4}
\vspace{-0.01in}
\end{figure}
\subsubsection{CRACK versus NR-RIS Block Size}
In Fig.~\ref{p4}, we explore CRACK performance when the NR-RIS employs connections with different block sizes $L$. Since larger $L$ provide greater flexibility in creating non-reciprocal effects, the ergodic sum rate sharply decreases at first before stabilizing at a low level. We see that only a relatively small block size (e.g. $L=8$) is needed to achieve the same performance degradation as the largest possible block size (i.e. $L=N=128$). Choosing the smaller value for $L$ is obviously preferred due to its reduced implementation complexity.

\begin{figure}[!t]
\begin{center}
\vspace{-0.01in}
\includegraphics[width=3 in]{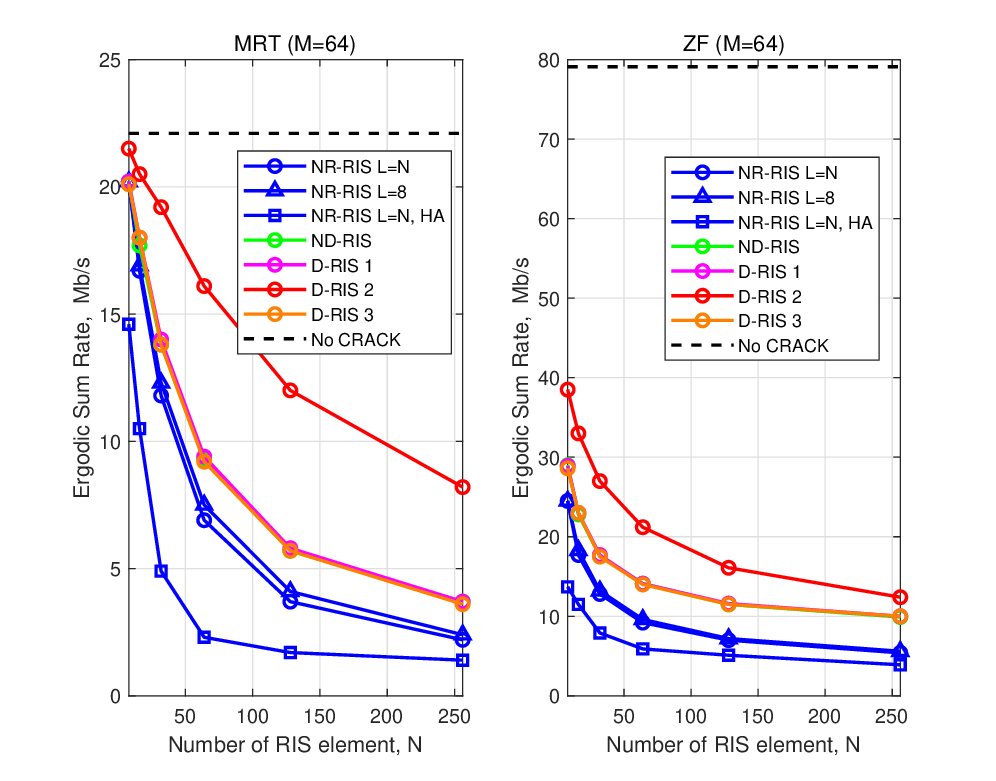}
\end{center}
\vspace{-0.1in}
\caption{Attack effect comparisons among NR-RIS, ND-RIS and D-RIS.}\label{p5}
\vspace{-0.1in}
\end{figure}

\subsubsection{Performance Comparison with Benchmarks}
This example contrasts the effectiveness of the physically consistent NR-RIS CRACK approach with the idealized ND-RIS CRACK model proposed in~\cite{10445725} and the three D-RIS-based attack strategies from~\cite{10081025,10149173,10424421}. As shown in Fig.~\ref{p5}, increasing the RIS size $N$ leads to a marked decrease in ergodic sum rate for all methods. As before, the physically consistent NR-RIS with small block size $L=8$ achieves performance comparable to the fully flexible case with $G=1$ and $L=N$, and both are superior to ND-RIS CRACK, D-RIS, and D-RIS 3 (the performance of these latter three are virtually indistinguishable on the plot). Among all the non-optimized strategies, NR-RIS CRACK induces the most severe degradation without requiring synchronization with the legitimate system. Finally we note that HA is able to achieve a significantly stronger attack by exploiting partial CSI for the legitimate system.

\subsection{CRACK aided Eavesdropping}
In this subsection, we investigate the security threats posed by an NR-RIS CRACK when a single-antenna eavesdropper intercepts information intended for legitimate users. As above, the NR-RIS is randomly configured and not optimized for the eavesdropper.
\begin{figure}[!t]
\begin{center}
\vspace{-0.10 in}
\hspace*{-0.5cm}\includegraphics[width=3 in]{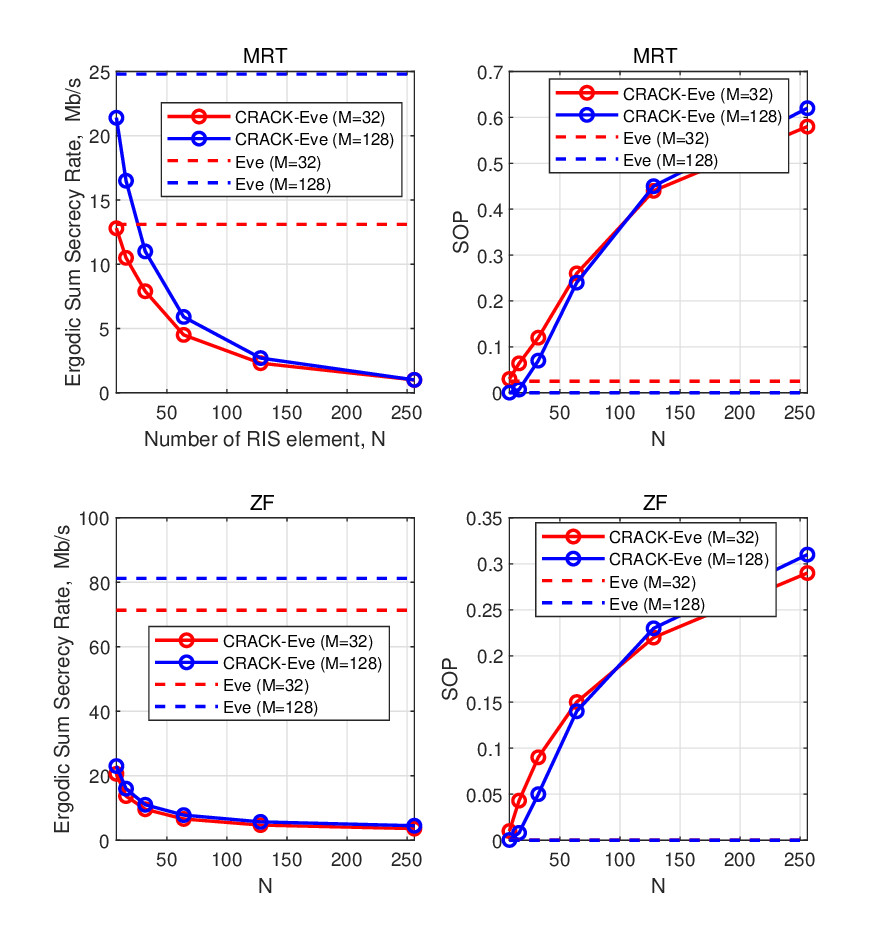}
\end{center}
\vspace{-0.2in}
\caption{Ergodic sum secrecy rate and SOP versus $N$.}\label{p6}
\end{figure}

\subsubsection{CRACK-Eve versus NR-RIS Size}
Fig.~\ref{p6} shows the ergodic sum secrecy rate and the corresponding SOP as functions of $N$ for $M=32$ and $M=128$ BS antennas. In the absence of CRACK, it is difficult for passive eavesdropping to succeed, particularly under ZF precoding, where the SOP is normally near zero since the signal energy is  focused only on the intended users with minimal leakage. However, CRACK disrupts this advantage by compromising the precoder design. As $N$ increases, the secrecy rate rapidly deteriorates and eventually stabilizes at a low level, even for large $M$. Concurrently, the SOP continues to rise. These results underscore the serious security threat CRACK poses to TDD systems, even when the NR-RIS operates with random configurations.

\begin{figure}[ht]
\begin{center}
\vspace{-0.1in}
\hspace*{-0.5cm}\includegraphics[width=3 in]{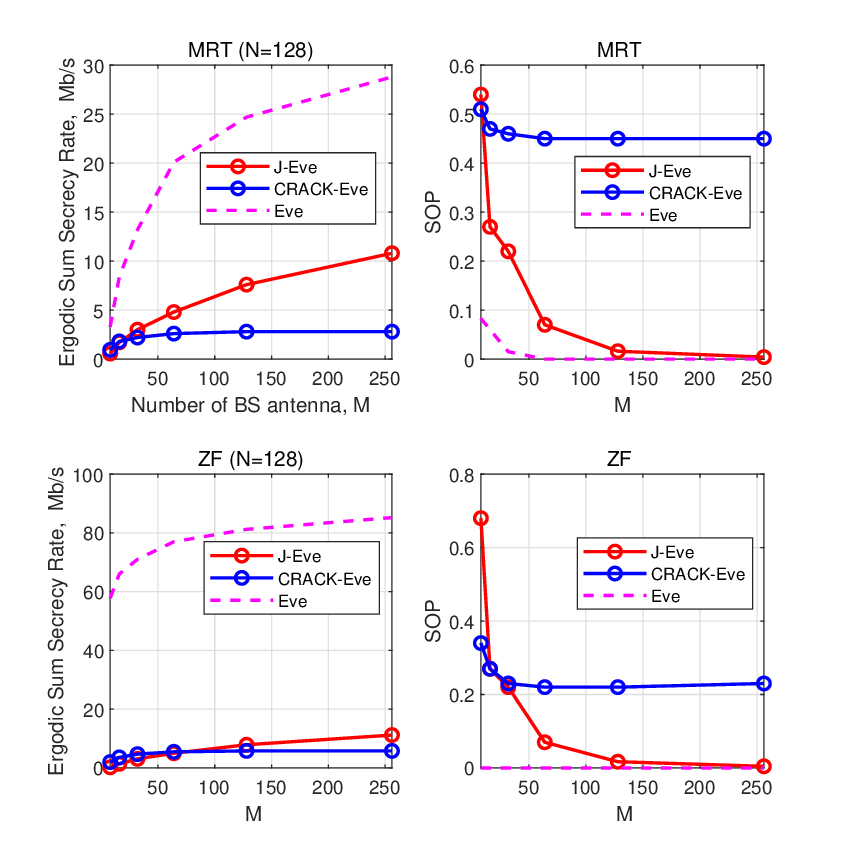}
\end{center}
\vspace{-0.2in}
\caption{Ergodic sum secrecy rate and SOP versus $M$.}\label{p7}
\end{figure}
\subsubsection{CRACK-Eve versus BS Antenna Array Size}
Fig.~\ref{p7} presents the ergodic sum secrecy rate and the corresponding SOP versus the number of BS antennas $M$. When $M$ is small, both J-Eve and CRACK-Eve significantly degrade the user rate, resulting in frequent secrecy outages. As $M$ increases, the impact of J-Eve is gradually mitigated; the secrecy rate under MRT improves, and the SOP for both MRT and ZF approaches zero, owing to enhanced beamforming directivity that suppresses signal leakage. However, this trend does not hold for CRACK-Eve, as the effect of mismatched precoding remains, indicating that increasing $M$ alone is insufficient to eliminate the security threat posed by CRACK.

\subsection{DRL-Based SecureCoder against Covert CRACK}
The following simulation results illustrate the performance enhancement achieved by the proposed SecureCoder algorithm when the legitimate TDD system is subject to covert CRACK and the corresponding information leakage threat.

\subsubsection{Convergence of SecureCoder DRL Training Process}
\begin{figure}[!t]
\begin{center}
\vspace{-0.01in}
\hspace*{-0.5cm}\includegraphics[width=2.9 in]{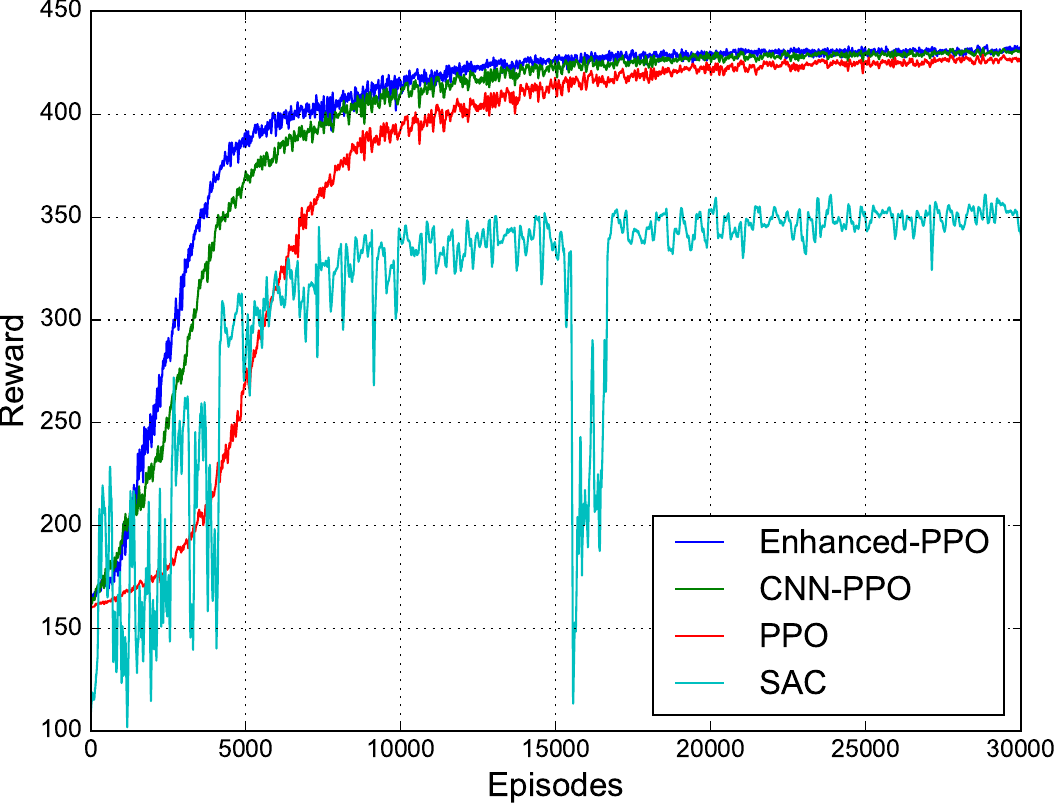}
\end{center}
\vspace{-0.05in}
\caption{Comparison of training convergence behavior.}\label{p8}
\vspace{-0.05in}
\end{figure}
Fig.~\ref{p8} presents the cumulative rewards per episode for our proposed enhanced-PPO algorithm, along with comparisons to three benchmark methods: standard PPO~\cite{schulman2017ppo}, CNN-PPO, and soft actor-critic (SAC)~\cite{haarnoja2018soft}. The results demonstrate that the PPO-based approaches exhibit superior stability and efficiency in dynamic wireless environments compared to SAC, primarily due to their use of clipped surrogate objectives that constrain policy updates within a bounded optimization space. The incorporation of CNN and PER further improves the agent's ability to develop effective transmission strategies when dealing with dynamic channels and sparse reward signals. Moreover, PPO's online-learning mechanism, which emphasizes recent interaction data during policy updates, enhances its application to practical communication scenarios.

\begin{figure}[!tbp]
\begin{center}
\vspace{-0.1in}
\includegraphics[width=3 in]{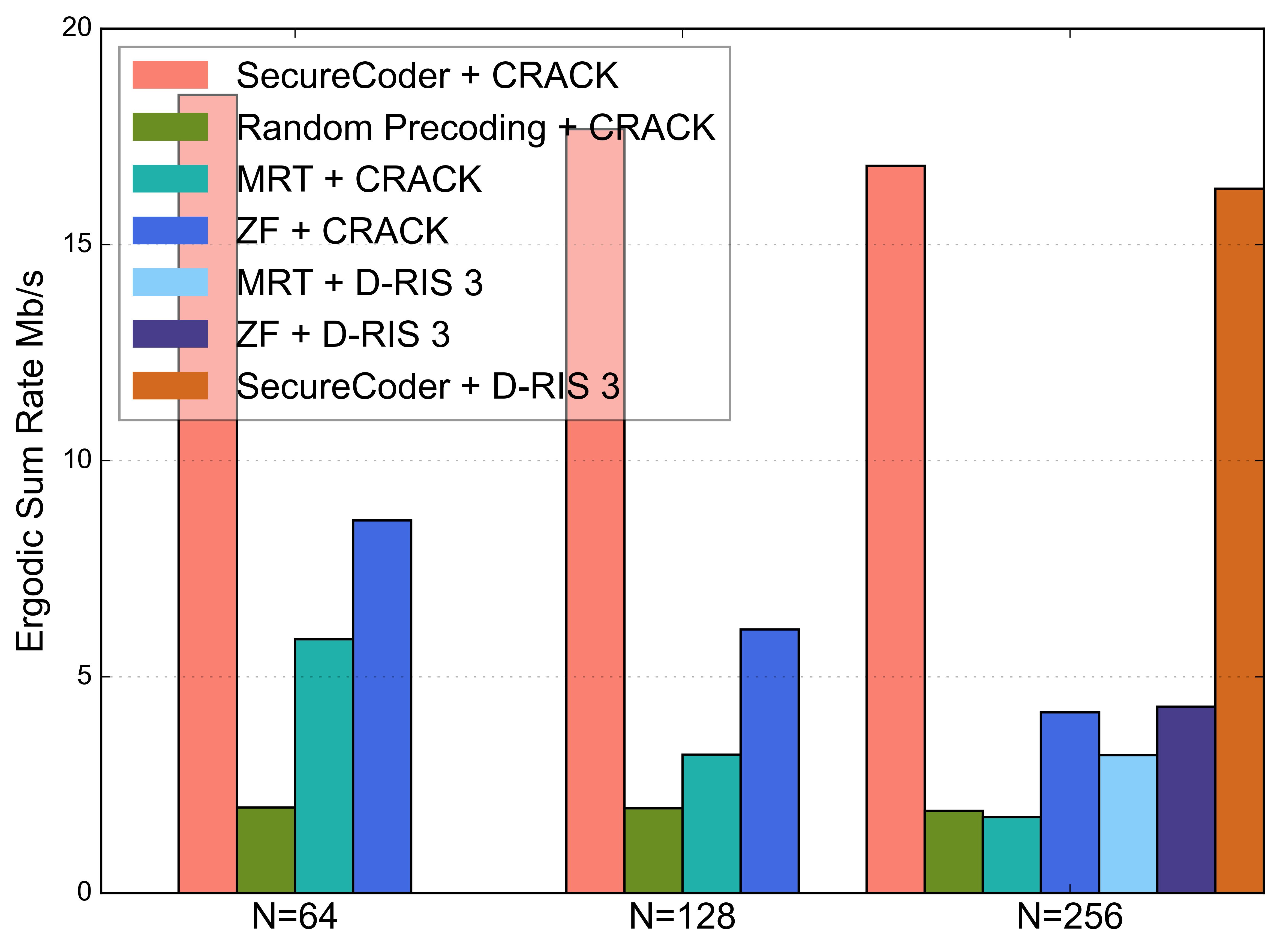}
\end{center}
\vspace{-0.05in}
\caption{Performance comparison for blind CRACK (M=32).}\label{p9}
\vspace{-0.05in}
\end{figure}
\subsubsection{SecureCoder and Blind CRACK} Fig.~\ref{p9} shows the ergodic sum rate achieved by SecureCoder and other precoding methods when the TDD system is subject to blind CRACK for NR-RIS with different number of elements. As $N$ increases, the disruption of the channel reciprocity leads to a significant decline in the performance of MRT and ZF precoding. In contrast, SecureCoder experiences only a slight decline and achieves performance superior to MRT without CRACK, although the gap with ideal ZF precoding remains quite large. Nonetheless, we observe that SecureCoder can achieve nearly a $300\%$ enhancement compared with the ergodic rate for ZF in a CRACK scenario with $N=256$. The simulation results further show that SecureCoder is effective in mitigating D-RIS-based ACA, yielding performance similar to the case of NR-RIS-based CRACK.

\begin{figure}[ht]
\centering
\subfigure[Ergodic sum secrecy rate]
{
    \begin{minipage}[b]{.8\linewidth}
        \centering
        \includegraphics[ width=2.9 in ]{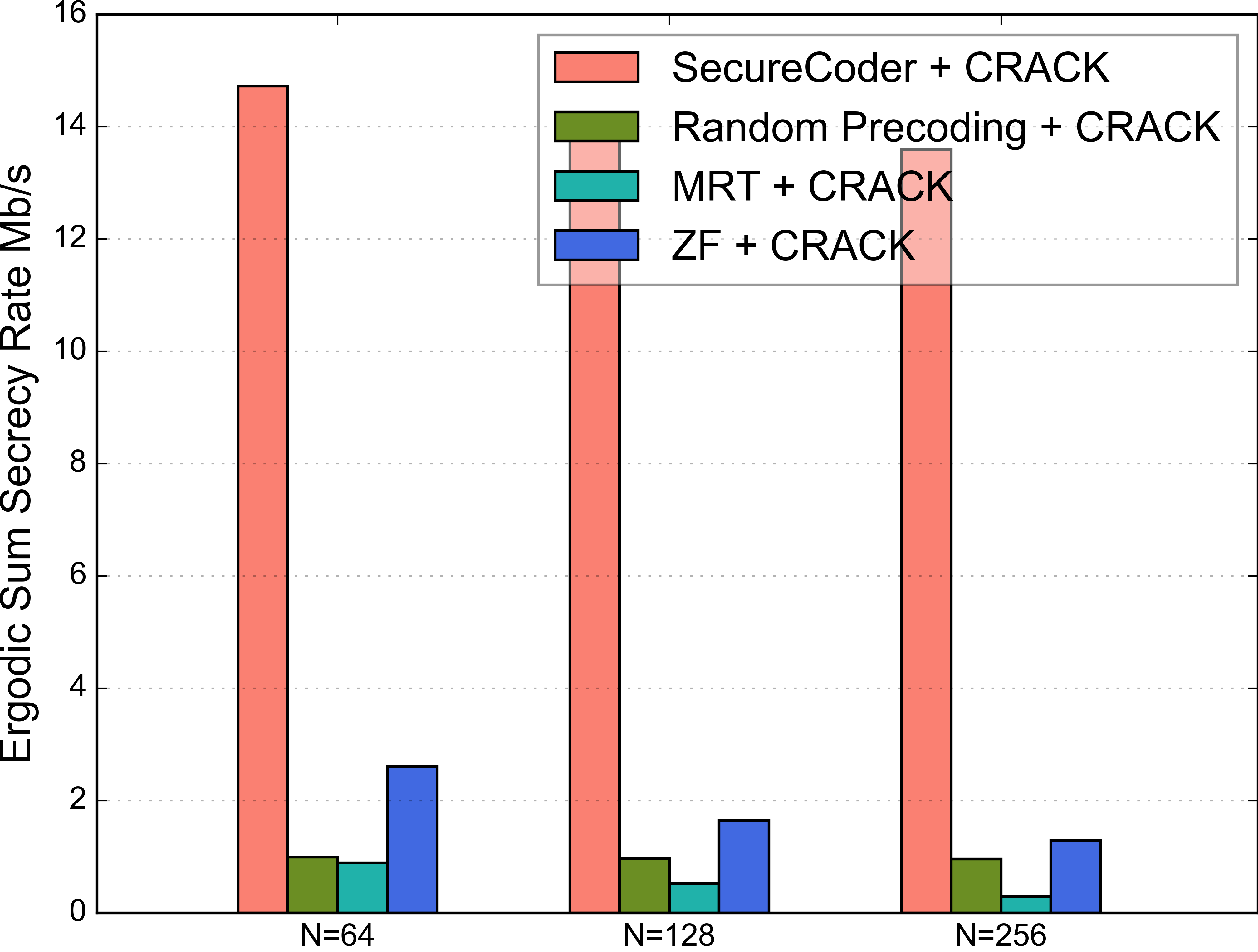}
    \end{minipage}
}

\subfigure[Secrecy Outage Probability]
{
 	\begin{minipage}[b]{.8\linewidth}
        \centering
        \includegraphics[width=2.9 in]{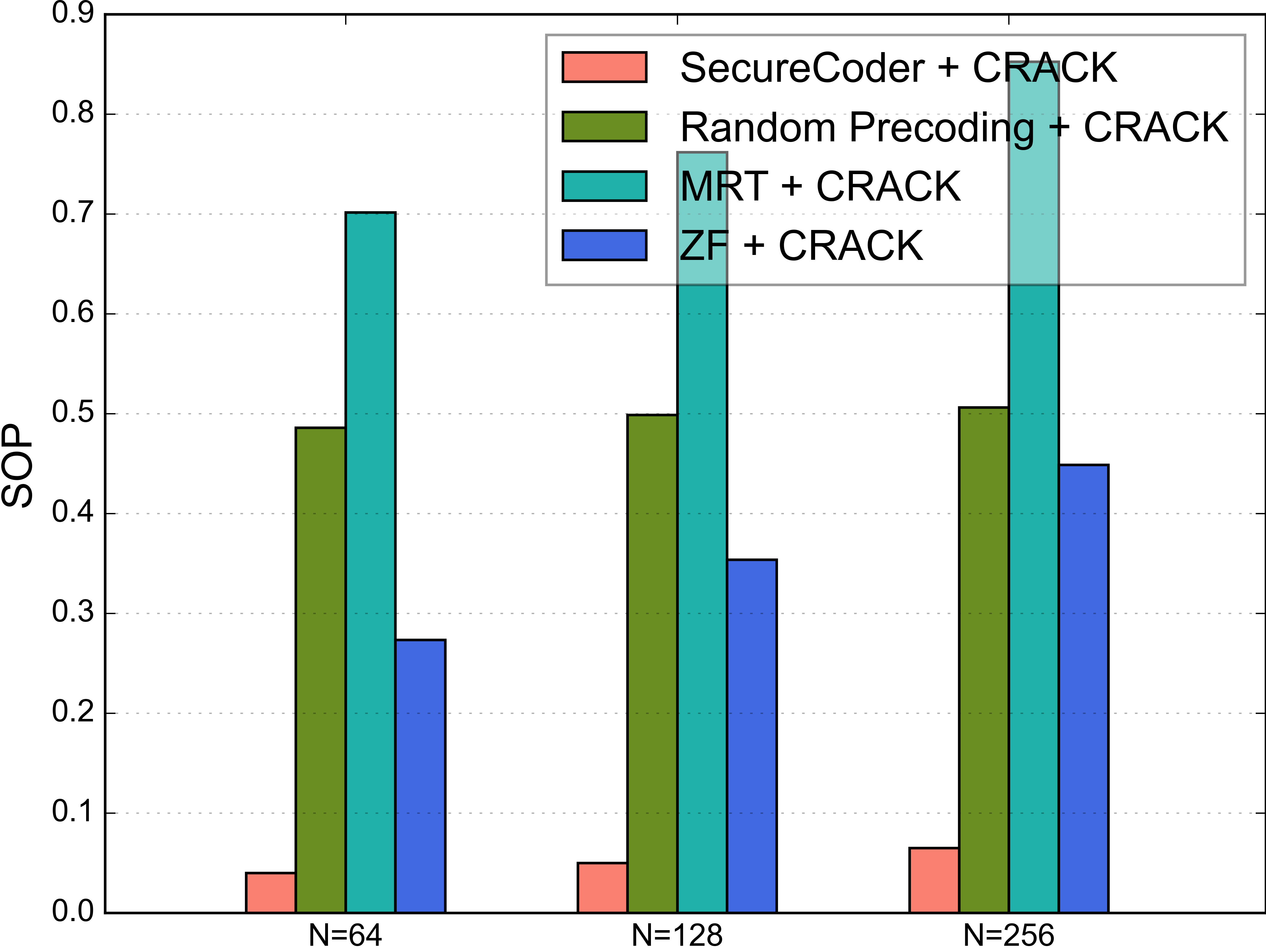}
    \end{minipage}
}
\vspace{-0.08in}
\caption{Performance comparison for CRACK with an eavesdropper (M=32).}\label{p10}
\vspace{-0.13in}
\end{figure}

\subsubsection{SecureCoder and Knowledge-driven CRACK-Eve} This example considers a scenario in which the adversary designs a set of NR-RIS configurations to implement CRACK-Eve, as in the scenario for Fig.~\ref{p1}. Fig.~\ref{p10} illustrates the ergodic sum secrecy rate and SOP achieved by SecureCoder and other precoding methods in this scenario. When $N$ is large (i.e. $N=256$), the BS is nearly unable to conduct confidential communication using either MRT and ZF precoding. In contrast, SecureCoder can enable a stable downlink secrecy rate while ensuring a low SOP. This is due to the design of the DRL reward function which not only guides the agent to learn a strategy that maximizes the sum rate, but prevents an uneven allocation of the rate among the users, making it difficult for Eve's rate to be superior to that of users with poorer channel conditions.

\section{Conclusions and Future Work}
In this paper, we presented a physically consistent NR-RIS model to examine the impact of a channel reciprocity attack (CRACK) in TDD communication systems. By passively disrupting the assumed channel reciprocity, CRACK introduces unforeseen multi-user interference that cannot be mitigated through conventional channel estimation or by simply increasing the number of base station (BS) antennas. Moreover, CRACK intensifies passive eavesdropping risks, resulting in frequent secrecy outages. Such an attack requires neither channel state information (CSI) nor synchronization with the pilot training or data transmission phases of the legitimate system. However, when knowledge of the statistical CSI or LoS paths is available, the impact of CRACK can be further amplified through targeted RIS design. To counter this threat, we developed SecureCoder, a DRL-based precoding approach that continuously interacts with the wireless environment. In this approach, the BS leverages estimated uplink CSI to adaptively generate downlink precoding, thereby restoring secure and stable transmissions.


\bibliographystyle{IEEEtran}
\bibliography{NRRIS_CRACK}

\begin{thebibliography}{10}
\providecommand{\url}[1]{#1}
\csname url@samestyle\endcsname
\providecommand{\newblock}{\relax}
\providecommand{\bibinfo}[2]{#2}
\providecommand{\BIBentrySTDinterwordspacing}{\spaceskip=0pt\relax}
\providecommand{\BIBentryALTinterwordstretchfactor}{4}
\providecommand{\BIBentryALTinterwordspacing}{\spaceskip=\fontdimen2\font plus
\BIBentryALTinterwordstretchfactor\fontdimen3\font minus \fontdimen4\font\relax}
\providecommand{\BIBforeignlanguage}[2]{{%
\expandafter\ifx\csname l@#1\endcsname\relax
\typeout{** WARNING: IEEEtran.bst: No hyphenation pattern has been}%
\typeout{** loaded for the language `#1'. Using the pattern for}%
\typeout{** the default language instead.}%
\else
\language=\csname l@#1\endcsname
\fi
#2}}
\providecommand{\BIBdecl}{\relax}
\BIBdecl

\bibitem{7467419}
Y.~{Zou}, J.~{Zhu}, X.~{Wang}, and L.~{Hanzo}, ``A survey on wireless security: Technical challenges, recent advances, and future trends,'' \emph{Proc. IEEE}, vol. 104, no.~9, pp. 1727--1765, Sept. 2016.

\bibitem{6739367}
A.~Mukherjee, S.~A.~A. Fakoorian, J.~Huang, and A.~L. Swindlehurst, ``Principles of physical layer security in multiuser wireless networks: A survey,'' \emph{IEEE Commun. Surv. $\&$ Tutor.}, vol.~16, no.~3, pp. 1550--1573, 3st Quat. 2014.

\bibitem{7081071}
N.~{Yang}, L.~{Wang}, G.~{Geraci}, M.~{Elkashlan}, J.~{Yuan}, and M.~D. {Renzo}, ``Safeguarding {5G} wireless communication networks using physical layer security,'' \emph{IEEE Commun. Mag.}, vol.~53, no.~4, pp. 20--27, Apr. 2015.

\bibitem{10409564}
R.~Kaur, B.~Bansal, S.~Majhi, S.~Jain, C.~Huang, and C.~Yuen, ``A survey on reconfigurable intelligent surface for physical layer security of next-generation wireless communications,'' \emph{IEEE Open J. Veh. Technol.}, vol.~5, pp. 172--199, Jan. 2024.

\bibitem{10188924}
W.~Khalid, M.~A.~U. Rehman, T.~Van~Chien, Z.~Kaleem, H.~Lee, and H.~Yu, ``Reconfigurable intelligent surface for physical layer security in {6G-IoT}: Designs, issues, and advances,'' \emph{IEEE Internet Things J.}, vol.~11, no.~2, pp. 3599--3613, Jul. 2024.

\bibitem{9475160}
C.~Pan, H.~Ren, K.~Wang, J.~F. Kolb, M.~Elkashlan, M.~Chen, M.~Di~Renzo, Y.~Hao, J.~Wang, A.~L. Swindlehurst, X.~You, and L.~Hanzo, ``Reconfigurable intelligent surfaces for {6G} systems: Principles, applications, and research directions,'' \emph{IEEE Commun. Mag.}, vol.~59, no.~6, pp. 14--20, June 2021.

\bibitem{9086766}
M.~A. ElMossallamy, H.~Zhang, L.~Song, K.~G. Seddik, Z.~Han, and G.~Y. Li, ``Reconfigurable intelligent surfaces for wireless communications: Principles, challenges, and opportunities,'' \emph{IEEE Trans. Cogn. Commun. Netw.}, vol.~6, no.~3, pp. 990--1002, Sept. 2020.

\bibitem{9424177}
Y.~Liu, X.~Liu, X.~Mu, T.~Hou, J.~Xu, M.~Di~Renzo, and N.~Al-Dhahir, ``Reconfigurable intelligent surfaces: Principles and opportunities,'' \emph{IEEE Commun. Surv. $\&$ Tutor.}, vol.~23, no.~3, pp. 1546--1577, 3st Quat. 2021.

\bibitem{9198898}
A.~Almohamad, A.~M. Tahir, A.~Al-Kababji, H.~M. Furqan, T.~Khattab, M.~O. Hasna, and H.~Arslan, ``Smart and secure wireless communications via reflecting intelligent surfaces: A short survey,'' \emph{IEEE Open J. Commun. Soc.}, vol.~1, pp. 1442--1456, Sept. 2020.

\bibitem{10143420}
M.~Hua, Q.~Wu, W.~Chen, O.~A. Dobre, and A.~L. Swindlehurst, ``Secure intelligent reflecting surface-aided integrated sensing and communication,'' \emph{IEEE Trans. Wireless Commun.}, vol.~23, no.~1, pp. 575--591, Jan. 2024.

\bibitem{9779086}
Y.~Sun, K.~An, Y.~Zhu, G.~Zheng, K.-K. Wong, S.~Chatzinotas, H.~Yin, and P.~Liu, ``{RIS}-assisted robust hybrid beamforming against simultaneous jamming and eavesdropping attacks,'' \emph{IEEE Trans. Wireless Commun.}, vol.~21, no.~11, pp. 9212--9231, Nov. 2022.

\bibitem{9439833}
J.~Zhang, H.~Du, Q.~Sun, B.~Ai, and D.~W.~K. Ng, ``Physical layer security enhancement with reconfigurable intelligent surface-aided networks,'' \emph{IEEE Trans. Inf. Forensics Secur.}, vol.~16, pp. 3480--3495, May 2021.

\bibitem{10143983}
G.~C. Alexandropoulos, K.~D. Katsanos, M.~Wen, and D.~B. Da~Costa, ``Counteracting eavesdropper attacks through reconfigurable intelligent surfaces: A new threat model and secrecy rate optimization,'' \emph{IEEE Open J. Commun. Soc.}, vol.~4, pp. 1285--1302, June 2023.

\bibitem{9501003}
G.~C. Alexandropoulos, K.~Katsanos, M.~Wen, and D.~B. Da~Costa, ``Safeguarding {MIMO} communications with reconfigurable metasurfaces and artificial noise,'' in \emph{Proc. IEEE Int'l Conf. on Communications (ICC)}, Montreal, Canada, June 2021.

\bibitem{9963962}
W.~Lv, J.~Bai, Q.~Yan, and H.~M. Wang, ``{RIS}-assisted green secure communications: Active {RIS} or passive {RIS}?'' \emph{IEEE Wireless Commun. Lett.}, vol.~12, no.~2, pp. 237--241, 2023.

\bibitem{9941040}
H.~Alakoca, M.~Namdar, S.~Aldirmaz-Colak, M.~Basaran, A.~Basgumus, L.~Durak-Ata, and H.~Yanikomeroglu, ``Metasurface manipulation attacks: Potential security threats of {RIS}-aided {6G} communications,'' \emph{IEEE Commun. Mag.}, vol.~61, no.~1, pp. 24--30, Jan. 2023.

\bibitem{wei2023metasurface}
M.~Wei, H.~Zhao, V.~Galdi, L.~Li, and T.~J. Cui, ``Metasurface-enabled smart wireless attacks at the physical layer,'' \emph{Nat Electron 6}, p. 610–618, Aug. 2023.

\bibitem{9789438}
Y.~Wang, H.~Lu, D.~Zhao, Y.~Deng, and A.~Nallanathan, ``Wireless communication in the presence of illegal reconfigurable intelligent surface: Signal leakage and interference attack,'' \emph{IEEE Wireless. Commun.}, vol.~29, no.~3, pp. 131--138, June 2022.

\bibitem{9605003}
S.~Sarp, H.~Tang, and Y.~Zhao, ``Use of intelligent reflecting surfaces for and against wireless communication security,'' in \emph{Proc. IEEE 4th 5G World Forum (5GWF)}, Montreal, Canada, Nov. 2021, pp. 374--377.

\bibitem{10073942}
L.~Dai, H.~Huang, C.~Zhang, and K.~Qiu, ``Silent flickering {RIS} aided covert attacks via intermittent cooperative jamming,'' \emph{IEEE Wireless Commun. Lett.}, vol.~12, no.~6, pp. 1027--1031, June 2023.

\bibitem{9112252}
B.~Lyu, D.~T. Hoang, S.~Gong, D.~Niyato, and D.~I. Kim, ``{IRS}-based wireless jamming attacks: When jammers can attack without power,'' \emph{IEEE Wireless Commun. Lett.}, vol.~9, no.~10, pp. 1663--1667, Oct. 2020.

\bibitem{10302337}
L.~Dai, C.~Zhang, S.~Wang, S.~Jia, H.~Huang, and K.~Qiu, ``Active {RIS}-empowered signal cancellation attacks,'' \emph{IEEE Trans. Vehic. Tech.}, vol.~73, no.~3, pp. 4487--4492, Mar. 2024.

\bibitem{10516473}
S.~Rivetti, O.~Demir, E.~Bj\"ornson, and M.~Skoglund, ``Malicious reconfigurable intelligent surfaces: How impactful can destructive beamforming be?'' \emph{IEEE Wireless Commun. Lett.}, vol.~13, no.~7, pp. 1918--1922, May 2024.

\bibitem{10081025}
H.~Huang, Y.~Zhang, H.~Zhang, C.~Zhang, and Z.~Han, ``Illegal intelligent reflecting surface based active channel aging: When jammer can attack without power and {CSI},'' \emph{IEEE Trans. Vehic. Tech.}, vol.~72, no.~8, pp. 11\,018--11\,022, Aug. 2023.

\bibitem{10149173}
H.~Huang, Y.~Zhang, H.~Zhang, Y.~Cai, A.~L. Swindlehurst, and Z.~Han, ``{Disco} intelligent reflecting surfaces: Active channel aging for fully-passive jamming attacks,'' \emph{IEEE Trans. Wireless Commun.}, vol.~23, no.~1, pp. 806--819, Jan. 2024.

\bibitem{10682037}
Y.~Zhang, H.~Huang, H.~Zhang, B.~Di, W.~Mei, J.~Yuan, and Y.~Cai, ``Disco intelligent omni-surface based fully-passive jamming attacks,'' in \emph{Proc. IEEE/CIC Int'l Conf. on Communications in China (ICCC)}, 2024, pp. 1881--1886.

\bibitem{10402016}
A.~S. de~Sena, J.~Kibiłda, N.~H. Mahmood, A.~Gomes, and M.~Latva-Aho, ``Malicious {RIS} versus massive {MIMO}: Securing multiple access against {RIS}-based jamming attacks,'' \emph{IEEE Wireless Commun. Lett.}, vol.~13, no.~4, pp. 989--993, Jan. 2024.

\bibitem{10424421}
H.~Huang, L.~Dai, H.~Zhang, Z.~Tian, Y.~Cai, C.~Zhang, A.~L. Swindlehurst, and Z.~Han, ``Anti-jamming precoding against {Disco} intelligent reflecting surfaces based fully-passive jamming attacks,'' \emph{IEEE Trans. Wireless Commun.}, vol.~23, no.~8, pp. 9315--9329, Feb. 2024.

\bibitem{10445725}
H.~Wang, Z.~Han, and A.~L. Swindlehurst, ``Channel reciprocity attacks using intelligent surfaces with non-diagonal phase shifts,'' \emph{IEEE Open J. Commun. Soc.}, vol.~5, pp. 1469--1485, 2024.

\bibitem{9737373}
Q.~Li, M.~El-Hajjar, I.~Hemadeh, A.~Shojaeifard, A.~A.~M. Mourad, B.~Clerckx, and L.~Hanzo, ``Reconfigurable intelligent surfaces relying on non-diagonal phase shift matrices,'' \emph{IEEE Trans. Vehic. Tech.}, vol.~71, no.~6, pp. 6367--6383, June 2022.

\bibitem{10302331}
J.~Singh, S.~Srivastava, A.~K. Jagannatham, and L.~Hanzo, ``Joint transceiver and reconfigurable intelligent surface design for multiuser {mmWave MIMO} systems relying on non-diagonal phase shift matrices,'' \emph{IEEE Open J. Commun. Society}, vol.~4, pp. 2897--2912, Oct. 2023.

\bibitem{10500386}
G.~Li, P.~Staat, H.~Li, M.~Heinrichs, C.~Zenger, R.~Kronberger, H.~Elders-Boll, C.~Paar, and A.~Hu, ``{RIS}-jamming: Breaking key consistency in channel reciprocity-based key generation,'' \emph{IEEE Trans. Info. Forensics $\&$ Security}, vol.~19, pp. 5090--5105, Apr. 2024.

\bibitem{10247269}
Z.~Wan, X.~Hu, X.~Sun, X.~Xu, K.~Huang, and L.~Jin, ``A countermeasure against {RIS} jamming attack in physical-layer key generation,'' \emph{IEEE Wireless Commun. Lett.}, vol.~12, no.~12, pp. 2193--2197, Sept. 2023.

\bibitem{10694006}
H.~Wang, J.~Nossek, and A.~Swindlehurst, ``Beyond-diagonal {RIS} attacks on physical layer key generation,'' in \emph{Proc. IEEE Int'l Workshop on Signal Processing Advances in Wireless Communications (SPAWC)}, 2024, pp. 946--950.

\bibitem{nrris}
\BIBentryALTinterwordspacing
J.~Xu, H.~Wang, R.~Liu, J.~A. Nossek, and A.~L. Swindlehurst, ``Non-reciprocal reconfigurable intelligent surfaces,'' 2024. [Online]. Available: \url{https://arxiv.org/abs/2411.15617}
\BIBentrySTDinterwordspacing

\bibitem{li2024non}
H.~Li and B.~Clerckx, ``Non-reciprocal beyond diagonal {RIS}: Multiport network models and performance benefits in full-duplex systems,'' \emph{arXiv preprint arXiv:2411.04370}, 2024.

\bibitem{10551389}
J.~A. Nossek, D.~Semmler, M.~Joham, and W.~Utschick, ``Physically consistent modeling of wireless links with reconfigurable intelligent surfaces using multiport network analysis,'' \emph{IEEE Wireless Commun. Lett.}, vol.~13, no.~8, pp. 2240--2244, June 2024.

\bibitem{10453384}
M.~Nerini, S.~Shen, H.~Li, and B.~Clerckx, ``Beyond diagonal reconfigurable intelligent surfaces utilizing graph theory: Modeling, architecture design, and optimization,'' \emph{IEEE Trans. Wireless Commun.}, vol.~23, no.~8, pp. 9972--9985, Feb. 2024.

\bibitem{7321779}
J.~Xu, L.~Duan, and R.~Zhang, ``Proactive eavesdropping via jamming for rate maximization over rayleigh fading channels,'' \emph{IEEE Wireless Commun. Lett.}, vol.~5, no.~1, pp. 80--83, 2016.

\bibitem{9305288}
L.~Sun, Y.~Zhang, and A.~Swindlehurst, ``Alternate-jamming-aided wireless physical-layer surveillance: Protocol design and performance analysis,'' \emph{IEEE Trans. Info. Forensics $\&$ Security}, vol.~16, pp. 1989--2003, Dec. 2021.

\bibitem{6151778}
X.~Zhou, B.~Maham, and A.~Hjorungnes, ``Pilot contamination for active eavesdropping,'' \emph{IEEE Trans. Wireless Commun.}, vol.~11, no.~3, pp. 903--907, Feb. 2012.

\bibitem{9576644}
J.~Yang, X.~Ji, F.~Wang, K.~Huang, and L.~Guo, ``A novel pilot spoofing scheme via intelligent reflecting surface based on statistical {CSI},'' \emph{IEEE Trans. Veh. Technol.}, vol.~70, no.~12, pp. 12\,847--12\,857, Oct. 2021.

\bibitem{schulman2017ppo}
J.~Schulman, F.~Wolski, P.~Dhariwal, A.~Radford, and O.~Klimov, ``Proximal policy optimization algorithms,'' \emph{arXiv preprint arXiv:1707.06347}, 2017.

\bibitem{haarnoja2018soft}
T.~Haarnoja, A.~Zhou, P.~Abbeel, and S.~Levine, ``Soft actor-critic: Off-policy maximum entropy deep reinforcement learning with a stochastic actor,'' in \emph{Proc. Int'l Conf. on Machine Learning (ICML)}, 2018, pp. 1861--1870.

\end{thebibliography}

\end{document}